\documentclass[12pt,preprint]{aastex}

\begin{document}

\newcommand\etal{et al. }

\def\t0{\theta_{\circ}}
\def\muo{\mu_{\circ}}
\def\sd{\partial}
\def\be{\begin{equation}}
\def\en{\end{equation}}
\def\bv{\bf v}
\def\bvo{\bf v_{\circ}}
\def\ro{r_{\circ}}
\def\rhoo{\rho_{\circ}}
\def\etal{et al.\ }
\def\msun{M_{\sun}}
\def\rsun{R_{\sun}}
\def\lsun{L_{\sun}}
\def\msunyr{M_{\sun} \, yr^{-1}}
\def\kms{\rm \, km \, s^{-1}}
\def\mdot{\dot{M}}
\def\ha{H$\alpha \;$}
\def\ecs{\rm erg \, cm^{-2} \, s^{-1}}

\title{Disk evolution in the Ori OB1 association}
\author{Nuria Calvet \altaffilmark{1},
Cesar Brice\~no\altaffilmark{2},
Jesus Hern\'andez\altaffilmark{2},
Sergio Hoyer\altaffilmark{3},
Lee Hartmann\altaffilmark{1}, 
Aurora Sicilia-Aguilar\altaffilmark{1},
S. T. Megeath\altaffilmark{1}, and
Paola D'Alessio\altaffilmark{4}
}

\altaffiltext{1}{Smithsonian Astrophysical Observatory, Mail Stop 42, Cambridge, MA 02138, USA;
Electronic mail: ncalvet@cfa.harvard.edu,hartmann@cfa.harvard.edu}

\altaffiltext{2}{Centro de Investigaciones de Astronom{\'\i}a (CIDA),
    Apartado Postal 264, M\'erida 5101-A, Venezuela;
Electronic mail: briceno@cida.ve,jesush@cida.ve,avivas@cida.ve}

\altaffiltext{3}{Dept. Astronom{\'\i}a y Astrof{\'\i}sica,
        Pontificia Universidad Cat\'olica de Chile,
        Campus San Joaquin, Vicu\~na Mackenna 4860 Casilla 306
        Santiago 22, Chile}

\altaffiltext{4}{Centro de Radioastronom\'\i a y Astrof\'\i sica,
Ap.P. 72-3 (Xangari), 58089 Morelia, Michoac\'an, M\'exico}

\begin{abstract}
We analyze multi-band photometry of a subsample
of low mass stars in the associations
Ori OB1a and 1b discovered during
the CIDA Orion Variability Survey,
which have ages of 7 - 10 Myr and
3 - 5 Myr, respectively. We obtained $UBVR_cI_c$
photometry at Mt Hopkins for 6 Classical T Tauri stars (CTTS) and 26
Weak T Tauri stars (WTTS) in Ori OB1a, and for 21 CTTS and 2 WTTS in Ori OB1b.  
We also obtained $L$ band photometry for 14 CTTS at Mt. Hopkins, and 
10$\mu$m and 18$\mu$m photometry with OSCIR at Gemini
for 6 CTTS; of these, all 6 were detected
at 10$\mu$m while only one was detected at 18$\mu$m.
We estimate mass accretion rates from the
excess luminosity at $U$, and find that they
are consistent with determinations for a number
of other associations, with or without
high mass star formation. 
The observed decrease of mass accretion
rate with age is qualitatively consistent 
with predictions of
viscous evolution of accretion disks,
although other factors can also play a role
in slowing accretion rates. We compare
the excesses over photospheric fluxes
in $H-K$, and $K-L$, and $K-N$ with
the younger sample of Taurus and find an
overall decrease of disk emission 
from Taurus to Ori OB1b to Ori OB1a.
This decrease
implies that significant grain growth
and settling towards the midplane has taken place in 
the inner disks of Ori OB1. We compare the
SED of the star detected at both 10$\mu$m and 18$\mu$m
with disk models for similar stellar
and accretion parameters. We find
that the low $\le 18 \mu$m fluxes of this Ori OB1b
star cannot be due to the smaller disk
radius expected from viscous evolution in
the presence of the FUV radiation fields from the
OB stars in the association. Instead, 
we find that the disk of this star
is essentially a flat disk, with little
if any flaring, indicating a
a significant degree of dust 
settling towards the midplane, as
expected from dust evolution in protoplanetary disks.
\end{abstract}


\section{Introduction}

Most star formation studies are based on a few
nearby associations without massive stars,
as Taurus and Chameleon or in highly 
populated clusters with significant high
mass formation, like the Orion Nebula Cluster (ONC).
In recent years, several teams have begun
systematic studies of the populations of
the OB associations. These associations
cover large volumes in the sky and include
populations covering the entire mass range. 
OB associations have been subject of studies with Hipparcos
(Brown et al. 1997; de Zeeuw \etal 1999), which has helped pinpoint
their distances and mean ages. 
In a number of associations, the populations
display a gradient of ages, from
$\sim$ 1 Myr to a few $\sim$ 10 Myr, which 
has been interpreted as the result of triggered
star formation (Blaauw 1964). 
This range of ages makes the OB associations
the most suitable laboratory for studies of
protoplanetary disk evolution, because
it encompasses the time span in which
giant planets are expected to form (Pollack et al. 1996;
Alibert et al. 2004), and significant solid
evolution is expected to take place, as indicated by meteoritic
evidence (Podoseck \& Cassen 1996; Wood 2004).

We are carrying out a photometric and spectroscopic
survey of $\sim $ 128 degrees in the Orion OB1 association 
in order to identify the low mass population.  Initial results of this survey
have been presented by Brice\~no et al. (2001, 2004 = Paper I).
In Paper I, we identified 197 new low mass members 
in the subassociations
Ori OB1a and 1b, spanning the mass range from
$\sim$ 0.2 to 1.4 $\msun$. We found ages of $\sim$ 3 - 5 Myr
for Ori OB1b and $\sim$ 7 - 10 Myr for Ori OB1a, confirming age determinations
based on the OB stars (Blaauw 1964;  Warren \& Hesser 1977, 1978; Brown et al. 1994).
We also found that Ori OB1b could be identified with
a population associated with a ring of low density
gas and dust, probably formed by a supernova event,
with most of the Classical T Tauri stars (i.e.,
stars accreting from circumstellar disks; CTTS)
near the ring, similar to the case of $\lambda$ Ori
(Dolan \& Mathieu 1999, 2001, 2002). In contrast, Ori OB1a
is in a region devoid of gas and dust, in agreement
with its older age. We also found that the number
of CTTS decreases sharply between Ori OB1b and 1a, 
indicating the few disks remain by $\sim$ 10 Myr,
in agreement with findings from 
young clusters, some containing high mass stars,
and low mass associations
as TW Hya (Haisch, Lada, \& Lada 2003; 
Hillenbrand,
Carpenter, and Meyer 2004, in preparation;
Sicilia-Aguilar
et al. 2004, in preparation;
Muzerolle et al. 2000, 2001).

In this work, we present multiwavelength photometry
of a subset of the stars in Paper I, including stars in 
Ori OB1a, 1b and 1c. In particular, we have obtained
$U$ photometry of a significant number of CTTS
in these subassociations, 
from which we have determined
mass accretion rates for comparison
with disk evolution theories. We have
also obtained $L$ band and mid-infrared (IR) measurements
at 10$\mu$m and 18$\mu$m of a small subset of
stars in Ori OB1, which has allowed us to make assessments
of the state of the dust in their disks in comparison
with younger populations.
In \S 2 we present the photometry, and in \S 3, 
we examine the properties of the disks in
Ori OB1a and 1b in comparison with the Taurus populations.
We discuss the implications of these
results for disk evolution in \S \ref{sec_discussion}.

\section{Observations and Data Analysis}

The targets were selected
from the sample of the
Orion variability survey presented in Paper I,
which were originally found on the
basis of variability in the $V$ band using
the QuEST camera on the 1m Schmidt telescope
on the Venezuela National Observatory. 
Follow-up spectroscopy using the FAST Spectrograph (Fabricant
et al. 1998) on the 1.5m telescope of the Fred Lawrence Whipple 
Observatory (FLWO) at Mt. Hopkins, and the Hydra multi-fiber spectrograph
(Barden \& Armandroff 1995) on the WYIN/3.5m telescope at Kitt Peak,
confirmed their membership to Ori OB1 (cf. Paper I).

\subsection{Optical photometry}

We used the 4SHOOTER CCD array on the 1.2m telescope of the 
Fred L. Whipple Observatory on Mt. Hopkins
during two observing runs to obtain
$UBVR_cI_c$ photometry of our list of stars. The first run was
during the nights of November 29 through December 4, 2002, 
although observations could be done only in the second half of the night of December 3 
and all night on December 4. The second run was during
October 15 - 20, 2003. All six nights were clear.
The seeing was between {1.2\arcsec } and {2.5\arcsec } throughout the observations.
The 4SHOOTER camera contains four $2048\times 2048$ Loral CCDs separated
by {45\arcsec } and arranged in a $2\times 2$ grid. After binning $2\times 2$
during readout, the plate scale was ${\rm 0.67\arcsec \> pixel^{-1}}$.
In order to achieve more uniform and consistent measurements, we
placed all the target standard stars on the same CCD detector (chip 3).

We obtained $BVR_cI_c$ photometry for 65 stars 
in Ori OB1a, 1b, and 1c,
and $U$-band photometry for 59 of these stars.
In particular, out of 36 stars in Ori OB1a,
we obtained $U$ photometry for all 6 CTTS
and 26 WTTS;
for the 27 stars in Ori OB1b, we obtained $U$ photometry 
for the 2 WTTS and 21 CTTS; finally, we obtained $U$ photometry for the
2 CTTS observed in Ori OB1c.

The basic processing of the FLWO data was done
with IRAF\footnote{IRAF is distributed by the National Optical Astronomy Observatories,
which are operated by the Association of Universities for Research
in Astronomy, Inc., under cooperative agreement with the National
Science Foundation.} routines in the standard way.
For both runs, the $U$-band data were flat-fielded using sky flats
taken at dusk/dawn. 
Instrumental magnitudes were obtained using the APPHOT package in IRAF.
Though Orion is at a moderately low galactic latitude $b\sim -20^{\circ}$, 
our fields are not crowded in the relevant magnitude range, 
so aperture photometry is adequate. We used an aperture
radius of 15 pixels for the source, $\sim 4\times$ the typical 3.5 pixel
FWHM of our observations, large enough to include
all the light from the standard stars as well as some observations made
under mediocre seeing.  We used a sky annulus with of 
15 and 20 pixels for the inner and outer radii, respectively. 
The instrumental magnitudes were calibrated in the Johnson $UBV$ and
Cousins $RI$ system with observations of Landolt (1992) standard fields
SA92, SA95, SA113 and PG0231+051 at various air masses every night; each field 
contains stars with a range of colors similar to our target objects.
Our overall photometric errors are dominated by the transformation
uncertainty, which are 0.05 mag in $U$ and 0.02 magnitudes 
for $BV(RI)_C$. The nightly variation in the zero points
was 0.15, 0.07, 0.06, 0.04, 0.05 magnitudes for $UBV(RI)_C$ respectively,
suggesting reasonably good photometric stability throughout both runs.

The resultant photometry
is listed in Table \ref{tab_photometry}, where
stars are identified by their running numbers 
in Paper I. Two exposures in some bands were obtained
for star CVSO 35 in Ori OB1a and for stars CVSO 146 and 190 in Ori OB1b,
as indicated in Table \ref{tab_photometry}.

\subsection{Near-infrared photometry}

We obtained $L$-band photometry for a subsample
of 14 stars using the dual channel
infrared camera STELIRCam (Tollestrup \& Willner 1998)
on the 1.2m telescope at FLWO
during the nights of December 18-19, 2002.
STELIRCam consists of two $256\times 256$ pixel InSb detector arrays.
Each is fed from a dichroic mirror that separates
wavelengths longer and shorter than $1.9\mu$m into two independent
imaging channels (red and blue channels) for simultaneous observations
on the sky. 
Three separate magnifications can be selected by rotating
cold lens assemblies into the beam.
The magnification is the same in both wavelength channels. 
For our observations we used the medium lens 
yielding a scale of $\rm 0.6 pixel^{-1}$ and a $2.'5 \times 2.'5$ field
of view, together with the L-Barr filter ($3.5\mu$m) in the red channel.
To convert the instrumental magnitudes of STELIRCam  to
the standard system, stars BD+0.1694, G77-31,
Gl 105.5, Gl 406, HD 1160, HD 18881, HD 225023 and HD 40335 from the list
of Elias et al. (1982) were observed each night at various air masses.
Both the Orion and standard stars
were observed with a $3 \times 3$ square pattern with 20''
dither between positions. The integration time at each dither position
was 10 s ($0.1 \times 100$ coadditions).

All data were first linearized using frames obtained at various
exposure times and routines developed in the Interactive Data Language (IDL)
by T. Megeath at CfA.
We then used standard IRAF routines to proceed
with the data reduction. Average dark frames were constructed from darks
taken at the beginning and end of each night's observations. These average
dark frames were then subtracted from each image. Sky frames were individually
made for each observation by median-combining the nine unregistered frames
in one dither pattern. The sky frames were subtracted from each image within
the dither set.
The high and low airmass frames were grouped into pairs, and the low
airmass frame in each pair was subtracted from the high airmass frame.
These subtracted pairs were then normalized and the pairs combined using a
median statistic to create the flat field.

Typically only the target star was visible in the $L$-band images, with
a FWHM ranging from $\sim 2.3$ to 3 pixels. Thus, we measured instrumental
magnitudes with the IRAF APPHOT package, using an aperture radius
of 6 pixels, and a sky annulus with inner and outer radii of
10 and 20 pixels respectively, for both Orion and standard stars.
After obtaining the zero points and color terms (which were small)
we arrived at an $L$-band $1\sigma$ error = 0.07 mags, mostly dominated by
measurement errors.

The $L$-band photometry is presented in Table \ref{tab_photometry}, where we
also include 2MASS $JHK_s$ magnitudes for the
stars from Paper I.

\subsection{Mid-infrared photometry}

During the nights of December 3, 5 and 9, 2003
we obtained 10$\mu$m and 18$\mu$m photometry
of a subsample of 6 stars in Ori OB1a and B.
This subsample was selected among those CTTS with
$K_s < 11$ and with large excesses in the $JHK_s$ diagram.
The observations were obtained in service mode using OSCIR
on Gemini North at Mauna Kea.
OSCIR is a mid-infrared ($8-25\mu$m) imager and low/medium-resolution 
(R=100-1000) spectrograph,
which uses a $128 \times 128$ pixel Si:As IBC detector optimized 
for the wavelength range $8-25\mu$m. On Gemini, OSCIR has a plate 
scale of about 0.084 arcsec/pixel and a total field of view in its 
imaging configuration of $11\arcsec \times 11\arcsec$. 
We used the  N-wide filter
($\lambda_{eff}=10.8 \mu$m and $\Delta \lambda = 5.23 \mu$m) and the
IHW18 filter ($\lambda_{eff}=18.2 \mu$m and $\Delta \lambda = 1.65 \mu$m).
In the rest of this article we will refer to these filters as the
$10\mu$m and $18\mu$m filters.
Typical exposure times for the Orion targets
were 60s at $10\mu$m and 300s at $18\mu$m. Sky conditions
were mostly clear through the observations.
All the observations were performed using the standard technique of
chopping and nodding, with a chop throw of 15'' in declination.
For calibration of the photometry the standard star $\alpha$ CMa (Sirius)
was observed on all nights at both wavelengths. Flux density estimates
for Sirius were calculated using the SEDs published by Cohen et al. (1999).
For Sirius, integration times were of the order of 6s in both $10\mu$m
and $18\mu$m.
The data were processed through the standard pipeline for OSCIR
data. Chop pairs are subtracted: beam A - beam B or beam B - beam A
for each set, depending on the nod position. Then these differences
are averaged to form a final image. Because we were dealing with
point sources, per recommendation of the Gemini staff no flat fielding
was performed. The OSCIR array is very flat over the small number of
pixels within the point spread function.  Experience shared between Gemini
and University of Florida is that the errors associated with the photometric
calibration ($\sim 10\%$) are much larger than any flat field correction.

The $10\mu$m and $18\mu$m images showed only one object per frame,
when detected at all. We used APPHOT
to obtain aperture photometry of all targets.
Images at $10\mu$m showed typical FWHM$\sim 6$ pixels (or $\sim 0.5$
arcsec), so we used an aperture radius of 15 pixels and inner and outer
sky annulus radii of 15 and 25 pixels respectively.
Because most of the Orion stars were not detected at $18\mu$m,
we used consecutive pairs of $10\mu$m and $18\mu$m $\alpha$ CMa images
to determine average offset positions between the two filters,
and used these to place the aperture at the predicted position
in the $18\mu$m images of the Orion stars.
In this way we measured upper limits to the flux in this band.

The mid-IR measurements are listed in Table \ref{tab_oscir}.
All stars were detected at 10$\mu$m but only
one measurement was secured at 18$\mu$m.
We list 3-$\sigma$ upper limits in Table
2 for the non detections.

\section{Results}

\subsection{Environment of the sample}

\begin{figure}
\includegraphics[scale=1.8,angle=270,keepaspectratio=true]{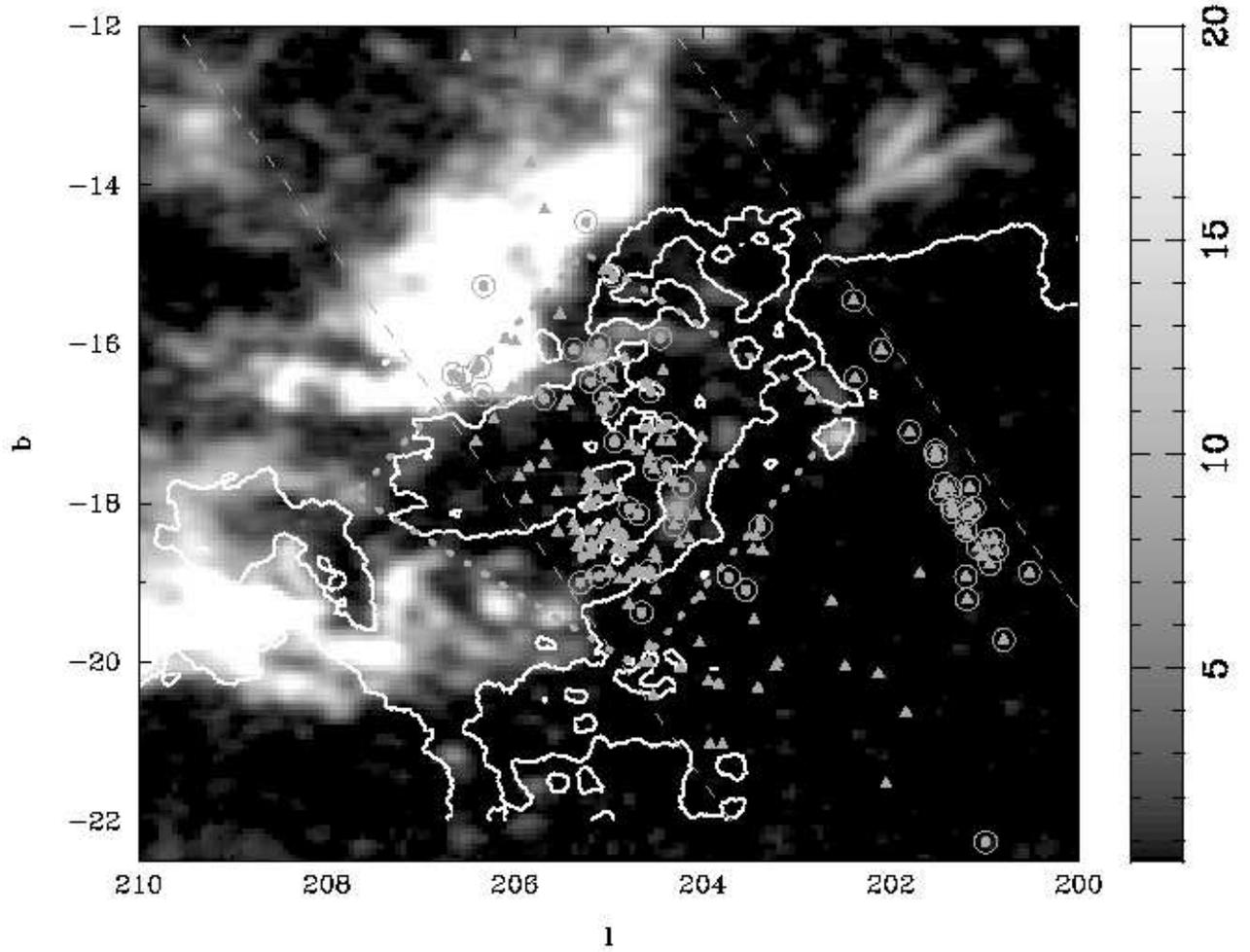}
\caption{Galactic coordinates of the stars in the photometric
sample (big open circles) compared to the Orion
Variability Survey sample of Paper I (CTTS, filled
circles, WTTS filled triangles). 
The integrated  $^{13}$CO emissivity
from Bally et al. (1987) is shown in
a halftone grey scale, covering the
range from 0.5 to 20 ${\rm K \, km \, s^{-1}}$.
The isocontour
for $A_V = $ 1 from Schlegel et al. (1998) is also
shown. The dotted lines show the
boundaries of Ori OB1b adopted in Paper I and
this work.
The dashed lines show the limits of the survey in
Paper I.
}
\label{fig_galactic}
\end{figure}

The galactic coordinates of stars in the photometric sample are plotted
in Figure \ref{fig_galactic} compared to the
entire sample in Paper I.
The stars are projected against the 
map of integrated  $^{13}$CO emissivity
from Bally et al. (1987).
It can be seen that our sample is representative
of that in Paper I. It includes the CTTS and
a number of WTTS 
in the ``clump'' 
at ($l = 202^o$,$b = - 18^o$) in Ori OB1a. It also includes a number of CTTS
in Ori OB1b, located mostly in the inside of
the molecular ring of higher extinction (cf. Paper I).
Some of these stars are probably 
associated with NGC 2023 and NGC 2024. 
Here we adopt the boundaries of Ori OB1b
from Paper I, shown in Figure \ref{fig_galactic}.
However, there is some uncertainty in the actual membership
of stars located near the assumed boundary between Ori OB1a and
1b at $l \sim 203.7^{\circ}$, $b \sim -18^{\circ}$ to $-19^{\circ}$.
In addition,
with the adopted boundaries, 2 stars of the sample fall
in Ori OB1c, in the region of high 
molecular gas density; however, they are so close to the
uncertain boundary that we include
them in Ori OB1b for the rest of the analysis.

\subsection{Stellar Properties}

We took the spectral types of the sample stars
from Paper I and recalculate here stellar properties
from the simultaneous multiwavelength photometry
to minimize effects of variability.
We used the average distances of
330 pc for Ori OB1a and 440 for Ori OB1b, as discussed in Paper I.
The extinctions $A_V$ in Table \ref{tab_prop} are calculated from
the $V-I_c$ color, using the Cardelli, Clayton,
\& Mathis (1989) extinction law with
$R_V$ = 3.1.
The mean extinctions for 1a, 1b, 1c, and the entire
sample are 0.5, 0.6, 2, and 0.6, respectively.
Using individual extinctions, we recalculated
stellar luminosities $L$ and radii $R$.
Location in the HR diagram and comparison
with the Siess et al. (2000) tracks yielded
stellar masses $M$ and ages. These quantities, shown in Table \ref{tab_prop},
in general agree within 50\% 
with those of Paper I.

\subsection{Color-color diagrams}
\label{sec_colcol}

\begin{figure}
\plotone{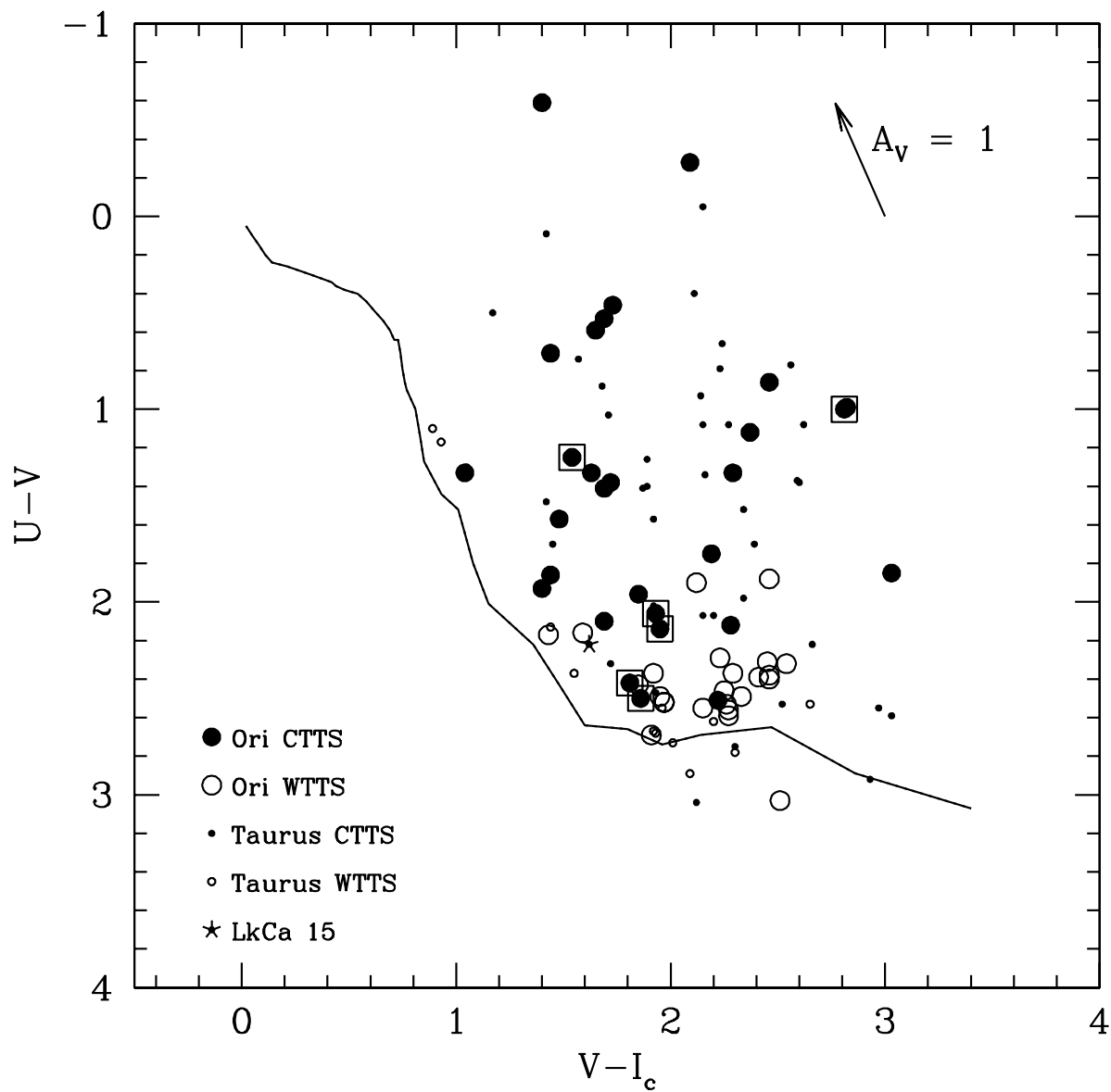}
\caption{Location of the stars in the $U-V$ vs. $V-I_c$
diagram. We show the CTTS (filled large circles), WTTS (open large
circles) for the Ori OB1 sample, with the stars
in Ori OB1a encircled by the open rectangle.
The Taurus CTTS (small filled circles) and WTTS
(small open circles),
and the sequence for dwarf
standards from Kenyon \& Hartmann (1995) are also 
shown,  as well as the
reddening vector for $A_V = 1$ calculated with
the Cardelli et al. (1989) reddening law
for $R_V = 3.1$. The 
star symbol corresponds to the Taurus CTTS LkCa 15.
}
\label{figviuv}
\end{figure}

\begin{figure}
\plotone{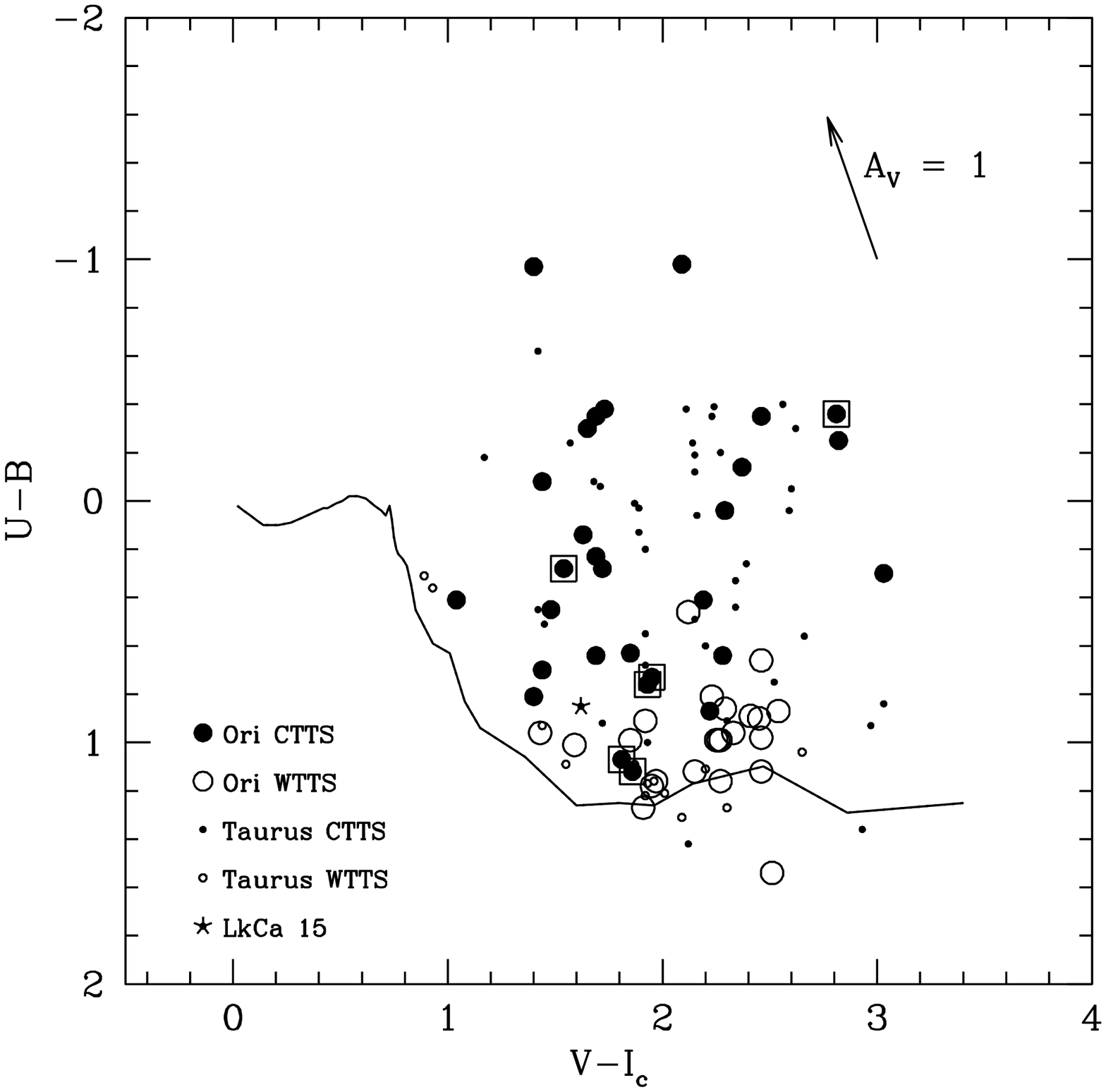}
\caption{Location of the stars in the $U-B$ vs. $V-I_c$
diagram. Symbols as in Figure \ref{figviuv}.
}
\label{figviub}
\end{figure}

Figures \ref{figviuv} and \ref{figviub} show 
photometry from Table \ref{tab_photometry}
plotted in the $U-V$ vs. $V-I_c$ and
$U-B$ vs. $V-I_c$ diagrams, as well as the
location of the main sequence colors
taken from Kenyon \& Hartmann (1995, KH95). 
CTTS and WTTS are shown separately, and the 
CTTS in Ori OB1a are marked. 
Note that the two observations for stars CVSO 35, 146, and 190
are plotted as different points.

The WTTS in the sample
are consistent with main sequence colors, reddened
by $\sim$ 0.5-0.6 magnitudes of extinction,
as expected from the average $A_V$ of the
sample. The CTTS have similar $V-I_c$ as the
WTTS, but  show large excesses relative to the
WTTS
in $U-V$ and $U-B$.
These excesses are expected
to arise from emission of the accretion shock
formed as material
accreting from the disk strikes the
stellar surface (Hartigan et al. 1991;
Hartigan, Edwards, \& Ghandour 1995;
Calvet \& Gullbring 1998).

In Figures \ref{figviuv} and \ref{figviub}
we also plot a sample of stars in Taurus taken from
KH95, equally separated in WTTS and
CTTS. 
The age of the Taurus association has been analyzed in detail by Hartmann
(2003), who showed that most of the stars have
ages of $\sim$ 1 - 2 Myr.
With the ages for the subassociations found in Paper I, i.e.
$\sim$ 3 - 5 Myr
for Ori OB1b and $\sim$ 7 - 10 Myr for OB1a,
the Taurus population is younger than Ori OB1b
by $\sim$ 2 - 3 Myr, and than Ori OB1a
by $\sim$ 6 - 8 Myr,
so it provides an initial reference point to study evolutionary effects.

The Orion OB1 and Taurus stars encompass a similar range
of $V-I_c$ in Figures \ref{figviuv} and \ref{figviub},
indicating that we are probing the same range
of stellar masses.
The samples of Ori OB1b and Taurus show
a similar range in $U-V$ and $U-B$ excesses
relative to photospheric colors;
the lowest excesses correspond to stars in Ori OB1a, 
some of them compatible with measurements
for the WTTS.

\begin{figure}
\plotone{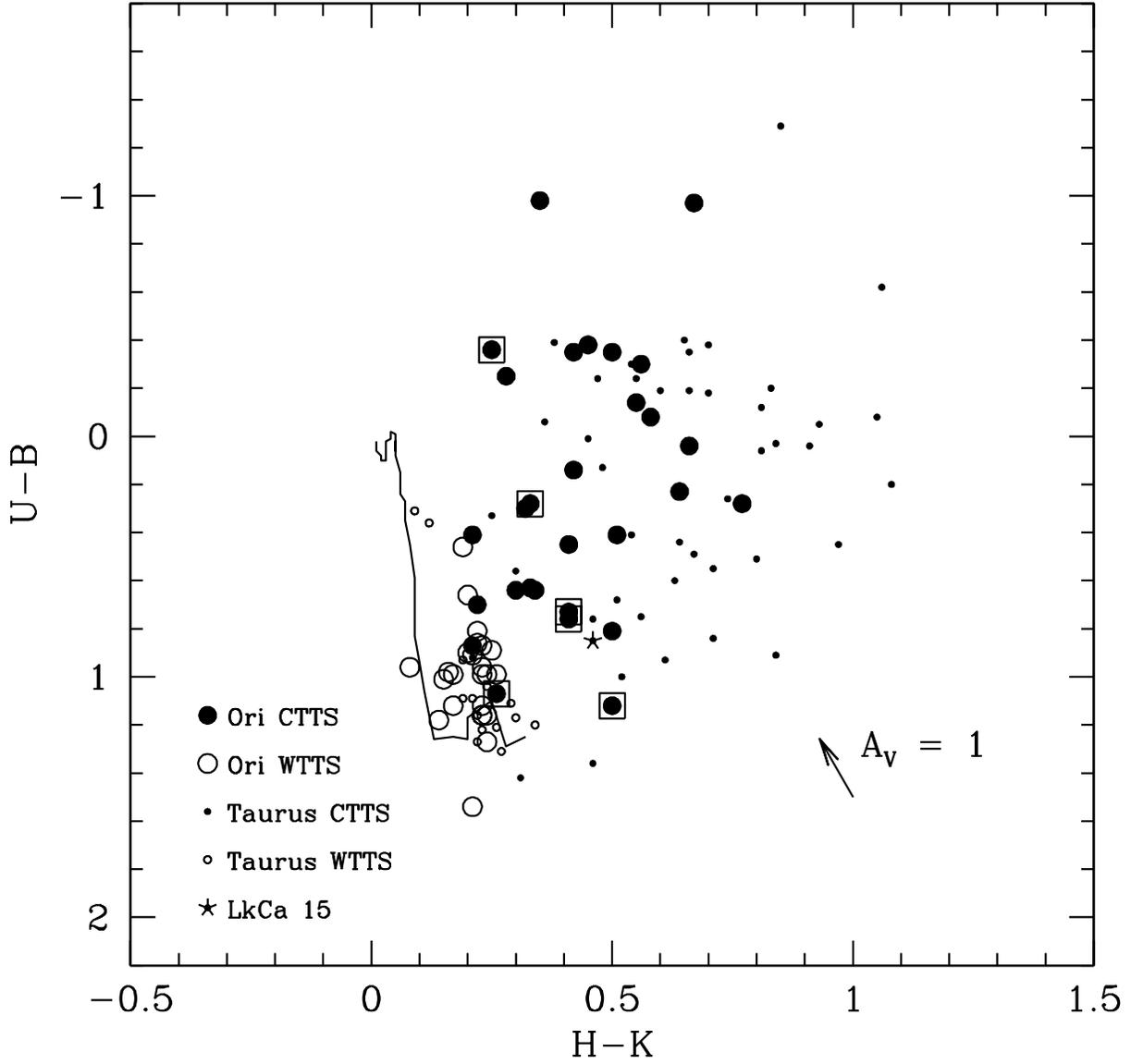}
\caption{Location of the stars in the $U-B$ vs. $H-K$
diagram. Symbols as in Figure \ref{figviuv}. The $H-K$ colors
for the Ori OB1 stars are from the 2MASS survey, corrected
to the CIT system.
}
\label{fighkub}
\end{figure}

In Figure \ref{fighkub} 
we show the Ori OB1 sample in the
$U-B$ vs. $H-K$ diagram, indicating 
the standard main sequence colors with
values from KH95. 
We have used the transformations
from Carpenter (2001) to convert
the 2MASS $H-K_s$ colors to the
CIT system.  A sample of Taurus stars,
with colors from KH95 is also shown.
It is apparent that the excesses in $H-K$
relative to photospheric colors
in Ori OB1b are smaller than those
found in the Taurus sample. Moreover,
the excesses in $H-K$ in Ori OB1a, the
oldest of the three populations shown,
are even smaller than in Ori OB1b. 

\begin{figure}
\plotone{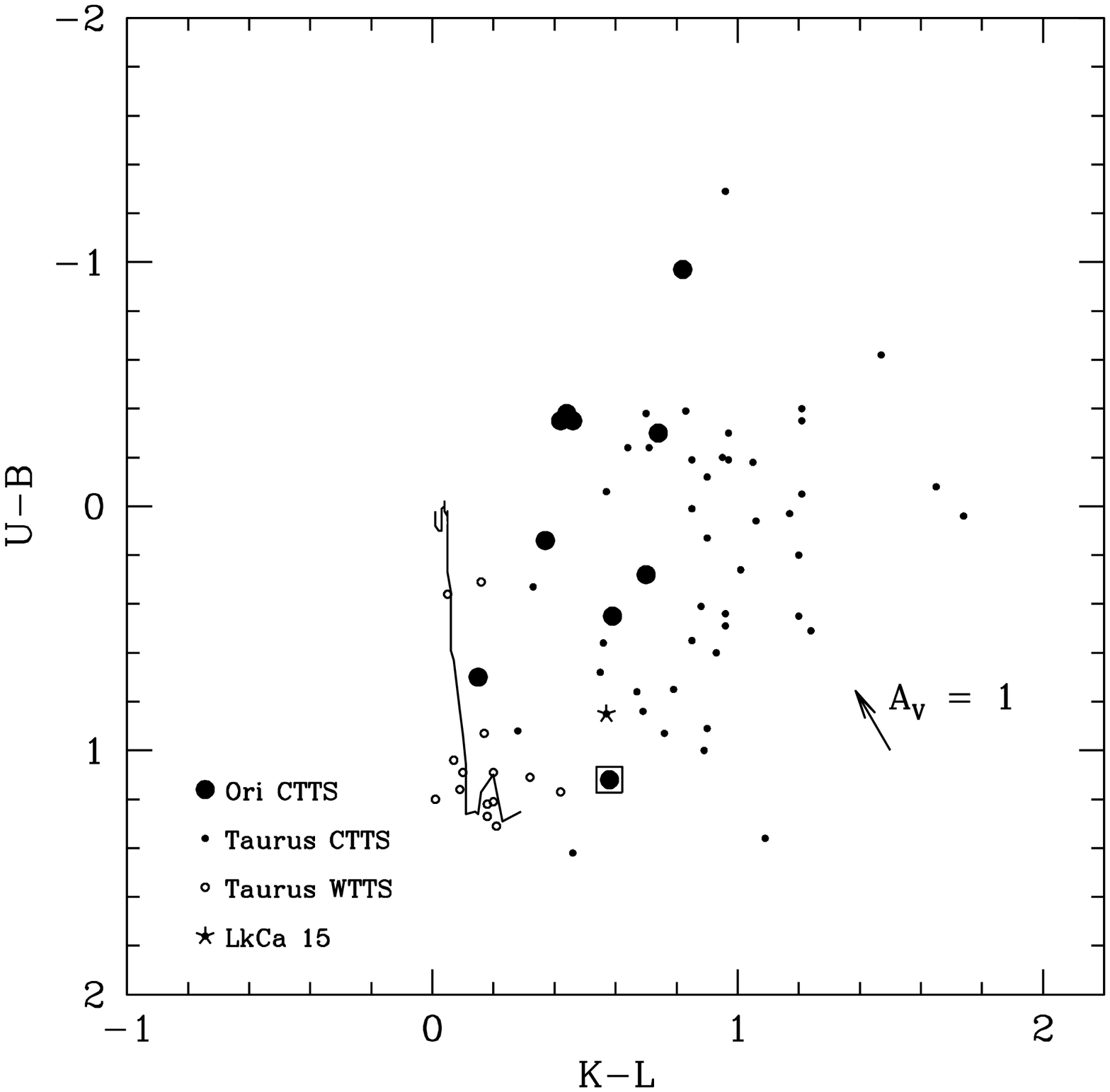}
\caption{Location of the stars in the $U-B$ vs. $K-L$
diagram. Symbols as in Figure \ref{figviuv}
}
\label{figklub}
\end{figure}

This effect is more apparent
in the $U-B$ vs. $K-L$ diagram,
shown in Figure \ref{figklub},
using the $K-L$ colors of the CTTS in Ori OB1 in Table \ref{tab_photometry}. 
We use here $K_s$ for the Orion OB1 stars.
Colors for the Taurus stars
and for the main sequence are from KH95.
The color excesses in $K-L$ relative to the
main sequence in the Ori OB1 stars
are clearly smaller than those in
Taurus. 

\begin{figure}
\plotone{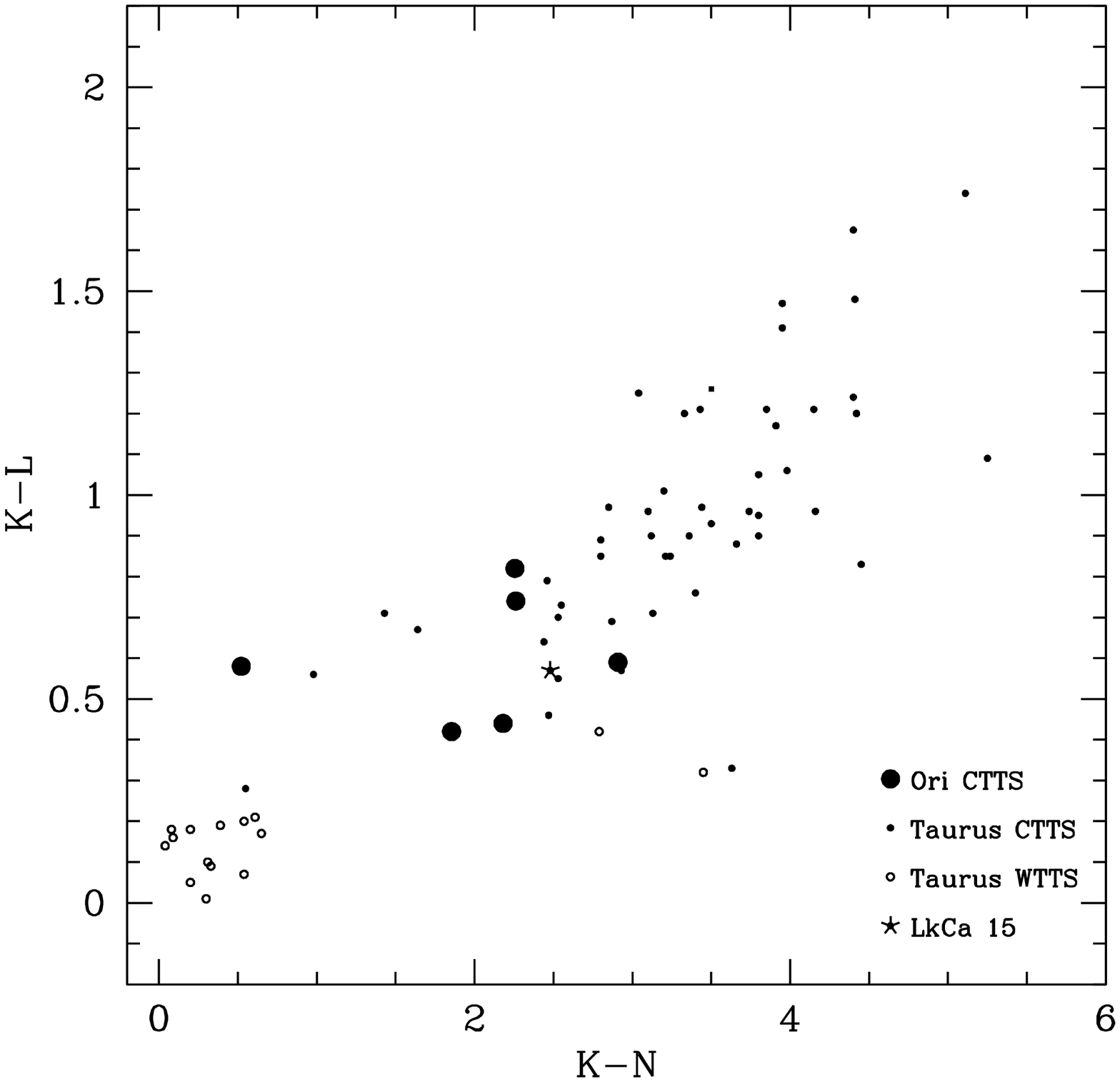}
\caption{Location of the stars in the $K-L$ vs. $K-N$
diagram. Symbols as in Figure \ref{figviuv}.
}
\label{figknkl}
\end{figure}

Finally, we locate the
6 stars observed with GEMINI/OSCIR
in the $K-N$ vs. $K-L$ diagram in Figure \ref{figknkl},
and compare it to the Taurus sample.
The Ori OB1 stars are located at the low end
of the Taurus CTTS distribution, and
even tend to populate the clear
gap between CTTS and TTS noticed
by KH95 (the three Taurus stars inside
the gap are binaries, KH95).

These comparisons clearly indicate 
a decrease of 
IR emission at $\le 10 \mu$m with age. 
A significant decrease of $H-K$
and $K-L$ with with age is also found
in a sample of clusters and 
groupings by Hillenbrand, Carpenter, \&
Meyer (2004, in preparation);
Sicilia-Aguilar et al. (2004, in preparation)
observe a similar near-IR flux
deficit relative to Taurus in Spitzer/IRAC
colors of the $\sim$ 5 Myr cluster Tr 37.
IR emission  at $\le 10 \mu$m arises in
the inner disks in T Tauri stars, $<$ few AU,
which are sufficiently hot to produce such 
emission (Meyer et al. 1997; D'Alessio et al. 1998; 1999; 2001).
Thus, the observed decrease of fluxes with age
must be related to physical
phenomena occurring in these inner regions of the disks
as a result of disk evolution.
We will explore 
possible implications of these results
in \S \ref{sec_discussion}.

\subsection{Spectral energy distributions}
\label{sec_seds}

\begin{figure}
\plotone{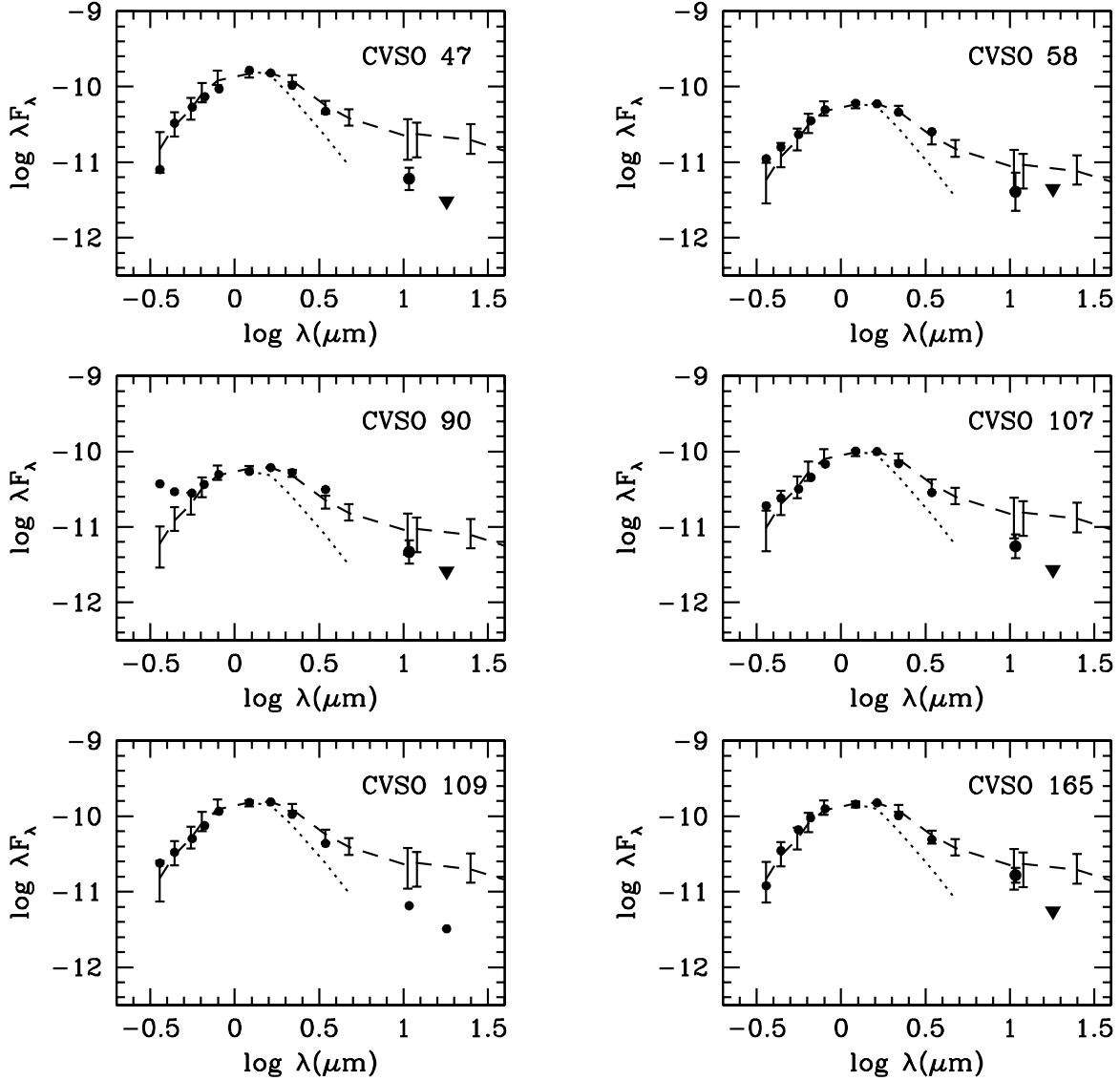}
\caption{Spectral energy distributions of stars observed
with GEMINI/OSCIR (filled circles) compared to the Taurus median
(D'Alessio et al. 2001, dashed lines),
normalized to the $H$
magnitude of each star. The error
bars in the Taurus median correspond to the first quartiles.
The inverted triangles indicate upper limits.
The photospheric fluxes in the near-IR  (dotted lines) are
constructed from colors of dwarfs from
Kenyon \& Hartmann (1995), normalized
to the $H$ magnitude of each star.
}
\label{fig_seds}
\end{figure}

Analyses of color-color diagrams in \S \ref{sec_colcol}
indicate a clear
difference between emission of the
disks in Ori OB1 and in Taurus; disks
in the Orion OB1 stars 
emit less flux at near-infrared wavelengths
than typical disks in Taurus.  The comparison is
seen clearly in Figure \ref{fig_seds}, where
we plot the spectral energy distributions (SEDs) of the stars
observed with Gemini OSCIR, including the
measurement and upper limits at 18 $\mu$m from Table \ref{tab_oscir}.
Optical and near-IR fluxes have been calculated from
magnitudes in Table \ref{tab_photometry} and corrected for
reddening in Table \ref{tab_prop}.  These SEDs are compared 
with the median SED of CTTS in Taurus (D'Alessio et al. 2001), scaled to
the flux at $H$ of each star.
The error bars in the median
show the first quartiles of the
distribution (i.e., 50\% of the Taurus
stars fall within the error bars).

The SEDs of the Orion stars show excesses relative
to the photosphere, consistent
with the fact that the
sample observed with
Gemini was selected
on the basis of being bright at $K$ and have
large $H-K$ excesses. However, even with this selection
bias, the Ori OB1 disks are fainter than $\sim$ 50 - 75 \%
of the Taurus sample in the mid-IR.

\subsection{Mass accretion rates}
\label{sec_mdots}

We interpret the excesses in $U-B$ and $U-V$
relative to photospheric fluxes in 
the CTTS in Ori OB1a and 1b as due
to emission from the accretion shock on
the stellar surface (Hartigan et al. 1991;
Hartigan, Edwards, \& Ghandour 1995;
Gullbring et al. 1998;
Calvet \& Gullbring 1998). 
In the framework of this interpretation,
the excess luminosity above photospheric
colors is a measure of the accretion
luminosity released as matter falls from
the disk onto the star, following
the magnetic field lines that disrupt the
disk (cf. Hartmann 1988).
We used the de-reddened $U$ photometry and spectral
types to obtain the excess luminosity in U
above the intrinsic photospheric fluxes, $L_U$;
this luminosity was used to estimate
the accretion luminosity $L_{acc}$,
using the calibration of Gullbring et al. (1998).
These estimates are listed in Table \ref{tab_prop}.
The largest uncertainty in the determination
of $L_U$, and thus $L_ {acc}$, comes from uncertainties in the
determination of the extinction $A_V$, and it amounts to $\sim 30 - 40$\%.

Using $L_{acc} \sim 0.8 G M \mdot /R_*$,
(assuming a radius of $\sim 5 R_*$ for the magnetospheric
radius; Calvet \& Gullbring 1998), with
the mass and radius determined from the location in the
HR diagram, we can infer the mass accretion
rate $\mdot$ of the objects. These rates
are listed in Table \ref{tab_prop}. 
In addition to the uncertainties in $L_{acc}$,
the formal uncertainty in $\mdot$
includes errors in 
radius and mass, which are of
order 15-20\%; thus, the 
mass accretion rates are constrained
within a factor of 2. Given the uncertainties
in the original calibration between $L_U$ and
$L_ {acc}$, this uncertainty can be a factor of 2
larger (Gullbring et al. 1998). 
As the errors are uncorrelated, the total
error amounts to a factor of 3.

\begin{figure}
\plotone{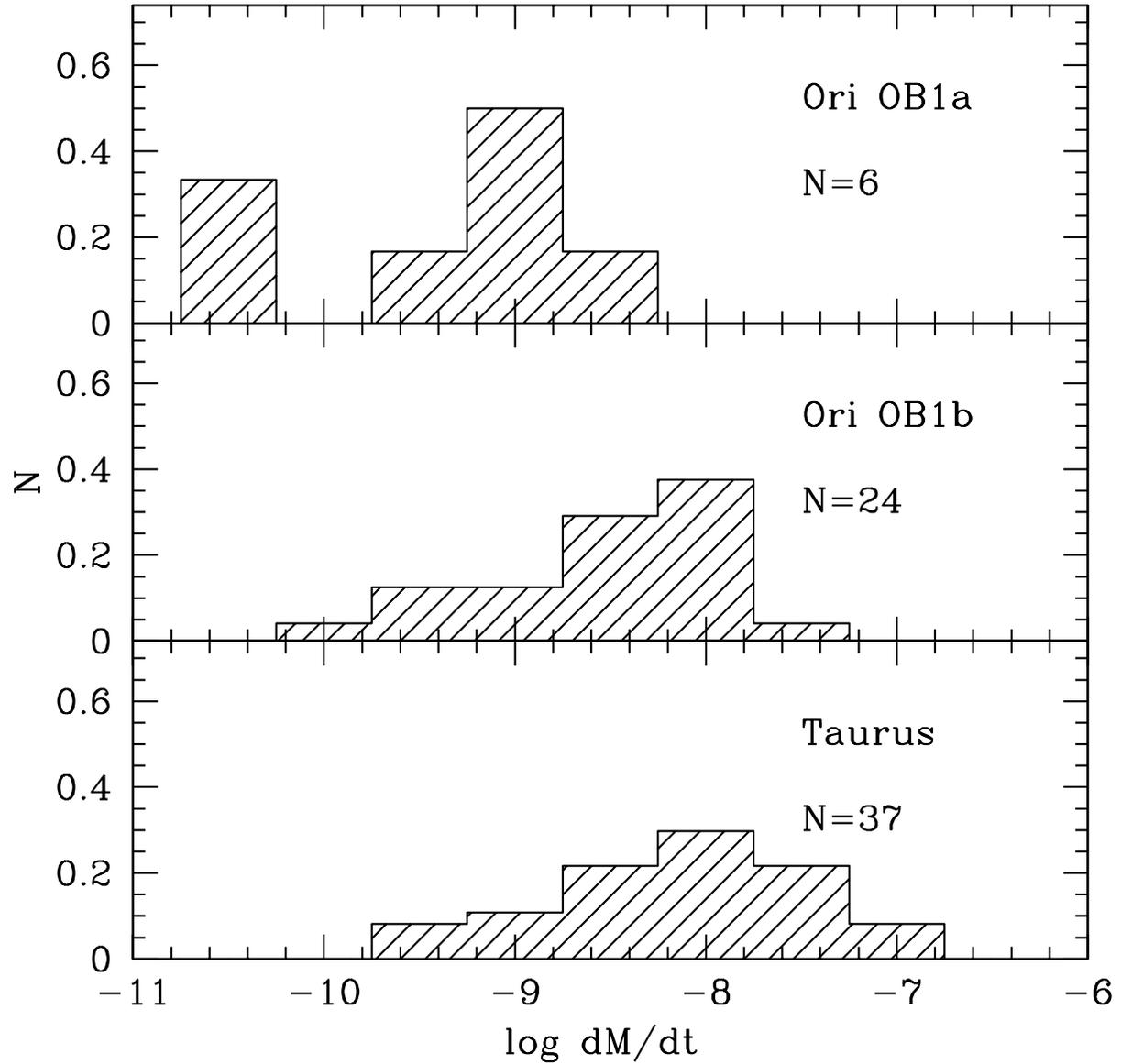}
\caption{Distribution of mass accretion
rate for Ori1 OB1a (upper panel), Ori OB1b (middle
panel) and Taurus (lower panel). Data from
Taurus from Gullbring et al. (1998) and
Hartmann et al. (1998), recalculated with masses
estimated from the Siess et al. (2000) evolutionary
tracks.
}
\label{fig_distmdot}
\end{figure}

In Figure \ref{fig_distmdot} we show histograms
with the number of objects per bin of log $\mdot$,
each of size 0.5, for the three populations,
Ori OB1a at the top, 1b at the middle, and
Taurus at the bottom. 
The mass accretion
rates in Taurus are taken from
Gullbring et al. (1998) and
Hartmann et al. (1998). 
These measurements of $\mdot$ were obtained from $L_{acc}$
using masses estimated with the D'Antona \& Mazitelli (1994)
evolutionary tracks;
we have recalculated them  using masses for the
Taurus stars estimated with the Siess et al. (2000)
evolutionary tracks, for consistency with the
work presented here.
The actual values of
$\mdot$ in bins with $\mdot < 10^{-10} \msunyr$
are highly uncertain. At these levels,
emission due to stellar surface
activity becomes an important or dominant
contributor to the U excess, as comparison
with the WTTS in the sample shows.
In these cases, the determined values are
upper limits to the true $\mdot$s.

The age of the three populations shown 
increases from bottom to top.
In all three cases, there is 
a large spread in the ranges of
values of $\mdot$.
However, as time proceeds, the number
of rapid accretors decreases,
only the low accretors
remain in 1a.
One ``continuum'' star is found in Ori OB1b
(see Table \ref{tab_prop}); we did not have
enough spectral resolution to 
distinguish spectral features and thus
determine
its spectral type and derived properties. 
This continuum star does not appear in 
Figure \ref{fig_distmdot}; if
its properties were similar to those of continuum stars
in Taurus, then it could have 
$\mdot \sim $ few $ 10^{-7} \msunyr$ (Gullbring et al. 1998;
Calvet \& Gullbring 1998; Gullbring et al. 2000).
However, even including this star, the relative number
of rapid accretors in Ori OB1b would be smaller than
in Taurus. 

We compared the distributions of log $\mdot$ in the three
regions using a two-sided KS test.  The results indicated
that the probability that the Ori OB1b and Taurus samples
are drawn from the same underlying distribution is only
3\%; the same is true for Ori OB1b vs. Ori OB1a.  
For a further test, we have added the continuum star in Ori OB1b
to the distribution, assuming a a mass accretion rate of
$10^{-7} \msunyr$; even in this case,
the probability
that the Ori OB1b sample is drawn from the same distribution
as Taurus is only marginally larger, 9\%.  Thus we conclude
that the distributions of accretion rates are significantly
different between the three regions.

\begin{figure}
\plotone{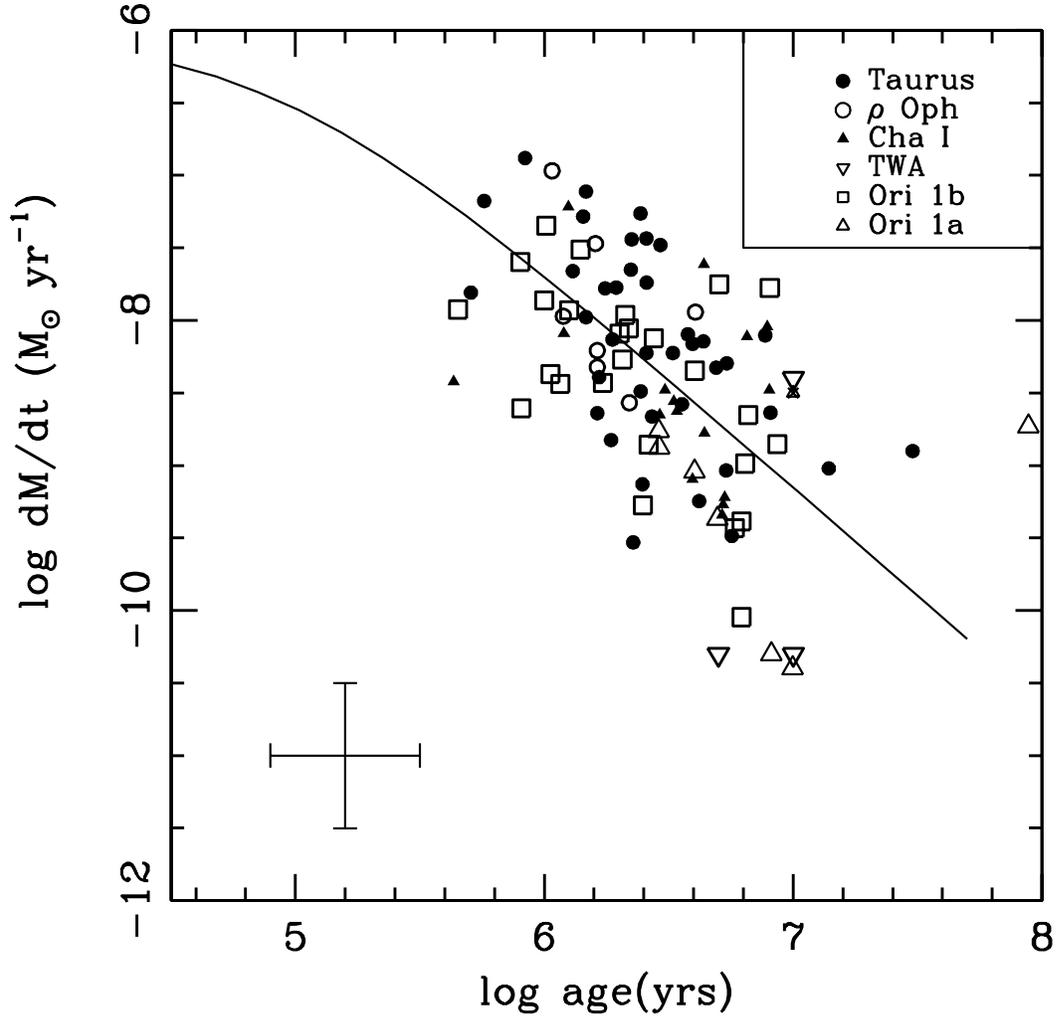}
\caption{Mass accretion rate vs. age for several
associations: Ori OB1a (open triangles),
Ori OB1b (open rectangles), Taurus (filled circles),
$\rho$ Oph (open circles), Chamaeleon I (filled
triangles), TWA (inverted open triangles), with
data from Hartmann et al. (1998) and Muzerolle et al.
(2000, 2001). Solid
line: model for viscous evolution for
an initial disk mass of $0.2 \msun$ and $\alpha = 0.01$
from Hartmann et al. (1998).
A typical error bar is shown.
}
\label{fig_mdot}
\end{figure}

Figure \ref{fig_mdot} plot the derived
mass accretion rates and ages on CTTS 
in Ori OB1a and 1b in Table \ref{tab_prop}. We also
plot in Figure \ref{fig_mdot} similar data for Taurus,
$\rho$ Ophiucus and Chameleon (Hartmann et al. 1998),
and the TW Hya association (TWA, Muzerolle et al.
2000, 2002). All masses and mass accretion
rates of these latter objects have
been re-calibrated using the Siess et al. (2000)
evolutionary tracks. 
The mass accretion rates of the disks in Ori OB1 are
consistent with those in other associations,
and show a clear decrease with age.
The star CVSO 41 in Ori OB1a has an apparent 
age of 88 Myr (Table \ref{tab_prop}) and
still shows a fairly high
value of $\mdot$. The spectral
type of this star is K2, so it could have
a large age uncertainty. It is
known that pre-main-sequence
stars in this spectral type range tend
to have higher apparent ages than cooler
objects (Hillenbrand 1997); this behavior has been attributed
to uncertainties in the birthline location (Hartmann 2003).

The solid line in Figure \ref{fig_mdot} shows 
the decay of $\mdot$ with time expected
from viscous evolution for a disk with initial mass $0.2 \msun$
and viscous parameter $\alpha = 0.01$, from
Hartmann et al. (1998). The
new values of $\mdot$ help confirm a decrease of $\mdot$ with age
qualitatively consistent with viscous disk evolution.
However, viscous evolution cannot explain
the decreasing fraction of accreting objects
with age (Muzerolle et al. 2000; Brice\~no et al. 2004);
other factors (i.e., inner disk clearing associated
with planet formation, Calvet et al. 2002; Rice et al. 2003;
D'Alessio et al. 2004) must play a role in slowing
accretion onto the central star.

\section{Discussion}
\label{sec_discussion}

In our comparison of disk properties in three
populations with ages $\sim$ 1-2 Myr (Taurus), 3-5 Myr (Ori OB1b), 
and 7-10 Myr (Ori OB1a),
we have found a clear decrease of disk
emission with age. This decrease is apparent
at all wavelengths with expected substantial
disk contribution to the emission, from $H$ to 18$\mu$m.

The disk emission in the IR depends on the temperature
of the dust. The two main heating mechanisms active on
disks around young stars are viscous dissipation and
stellar irradiation (Kenyon \& Hartmann 1987, KH87;
D'Alessio et al. 1998, 1999, 2001, D01). Viscous
dissipation depends on the mass accretion
rate throughout the disk; 
for the range of stellar parameters of stars
in Ori OB1,  and 
for $\mdot \le 10^{-8} \msunyr$, which characterize these
stars (\S \ref{sec_mdots}), viscous heating is important
only near the midplane of the innermost disk, $\le$ 1 AU
(D'Alessio et al. 1999, 2001). Thus, absorption
of stellar radiation by dust in the disks around the Ori OB1 stars
mostly determines their temperature and emission.

One important development in recent years
is the recognition that the near-IR fluxes in accretion
disks around pre-main sequence stars
are dominated by emission from
a sharp transition between dust and gas
at the dust destruction radius. At this
transition, the optically thick dust disk
is high enough to form a ``wall'', which
is frontally illuminated by the star.
Opacities in the inner gas disk are
due to molecules and are much lower
than the dust opacities for typically
low mas accretion rates; thus, the
gas disk is optically thin, enabling
frontal illumination of the wall (Muzerolle et al. 2004).
The existence of this wall was proposed for Herbig Ae/Be
stars (Natta et al. 2001; Tuthill, Monnier \& Danchi 2001;
Dullemond, Dominik, \& Natta et al. 2001)
to explain the peculiar ``bump'' at $\sim 3 \mu$m
observed in the SEDs of these stars (Hillenbrand
et al. 1992). Interferometric measurements
of inner disk structures in Herbig Ae/Be stars
confirmed this prediction (Millan-Gabet et al. 1999;
Monnier \& Millan-Gabet 2002).  A similar
wall emission was then reported in the low
mass CTTS by Muzerolle et al. (2003).
In contrast to their higher mass counterparts,
CTTS can have
accretion luminosities of the same
order as the stellar luminosities,
thus increasing the total energy output and
moving outwards the disk location where the temperature
is equal to the dust destruction temperature for objects with high values
of $L_{acc}$.
The rapid accretors are therefore expected to
have the wall located at larger radius
than objects with low $\mdot$, increasing the
total emitting area of the wall and thus 
the near-IR excess (Muzerolle et al. 2003; D'Alessio et al.  2003).
This prediction has been confirmed
by the Keck interferometer measurement
of an inner structure at the expected
location of the dust destruction radius in a high accretion
CTTS (Colavita et al. 2003; D'Alessio et al.  2003).

According to this model, the near-IR
excess depends on the strength of the
wall emission relative to photospheric fluxes. This emission is approximately
that of a black-body at the dust destruction
temperature, $T_d \sim $ 1400K (Muzerolle et al. 2003),
times the solid angle subtended by the wall.
This solid angle is essentially that
of a cylinder of given height, and radius
equal to the dust destruction radius,
with an inclination dependence which
makes the emission maximum at intermediate
lines of sight (60$^{\circ}$ - 80$^{\circ}$)
where the wall has maximum exposure (Dullemond
et al. 2001; D'Alessio et al.  2003).
The dust destruction radius
is essentially $R_d \propto (L_* + L_{acc})^{1/2} /T_d^2$
(D'Alessio et al.  2003). The height 
of the wall $z_w$ is estimated as the 
height where the radial optical depth
from the star becomes $\sim$ 1 (see Muzerolle et al. 2004). 

We can examine the decrease of near-IR excesses with age 
in terms of these ideas.
As they age, pre-main sequence low mass
stars move down along Hayashi tracks essentially
at constant $T_{eff}$ while decreasing in luminosity.
As $\mdot$ decreases, $L_{acc}$ becomes much smaller than  $L_*$,
so the dust destruction radius becomes fixed relative
to the stellar radius,
$R_d/R_* \propto L_*^{1/2} / T_d^2 R_* \sim (T_* / T_d)^2 \sim$ const.
If the height of the wall is a fixed
multiple of the disk scale height at $R_d$, which in
turn is typically
$\sim 0.1 R_d$, then the emitting area
of the wall, $\propto R_d z_w$, and thus its total energy output,
would decrease in the same proportion as
the stellar output, namely $\propto R_*^2$.
In this case, no decrease of near-IR excess
relative to the photosphere
with time would be expected. 
Since $R_d/R_* \sim$ const,
the most natural way to explain the observed
decrease is that the height of the
wall $z_w$ decreases with time. For this to happen,
the opacity at the upper levels of the 
wall has to decrease, which implies that
dust grains at the top of the wall significantly grow and/or
settle towards the midplane as age increases.

Grain growth and/or settling towards the
midplane in the disk regions outside the wall also result in 
low mid-IR fluxes.
The disk is heated by stellar + accretion energy
captured at the disk surface.
Thus, heating of the disk depends on the shape of this
surface; the more ``flared'' the surface is, the
more energy it can capture (Kenyon \& Hartmann 1987, KH87);
disks with highly flared surfaces are hotter
and thus emit more. 
To be more specific, the height of the surface $z_s$ at a given
radius is given by the condition that the radial optical
Since the density decreases with vertical height above the midplane,
$z_s$ decreases as the opacity of the dust 
in the upper layers decreases.
Small grains have high opacity at wavelengths characteristic
of the stellar radiation, in the optical 
and near-IR. They produce
a highly flared surface, which results
in high energy capture, and
hot and bright disks.
In contrast, large grains have
low opacity at short wavelengths. They produce
flatter surfaces, which result in less energy capture
and cooler and fainter disks (D'Alessio et al. 2001, D01).

Alternatively, the opacity at upper layers
may be reduced because grains are settling
to the midplane. This also results
in flatter surfaces and fainter disks in the infrared 
(Dullemond \& Dominik 2004; D'Alessio
et al. 2004, in preparation). 
Both growth
and settling are expected to occur together.
Dust particles, affected by gas drag, collide and stick together;
as the particles grow, they settle towards the midplane, falling under
the effect of gravity (Weidenschilling \& Cuzzi 1993;
Weidenschilling 1997). These theories have many
unknown parameters, and so far predict time scales
for complete settling $\le$ 1 Myr; however, the
timescales depend
upon the magnitude of turbulence in disks, which could
retard or prevent settling.

\begin{figure}
\plotone{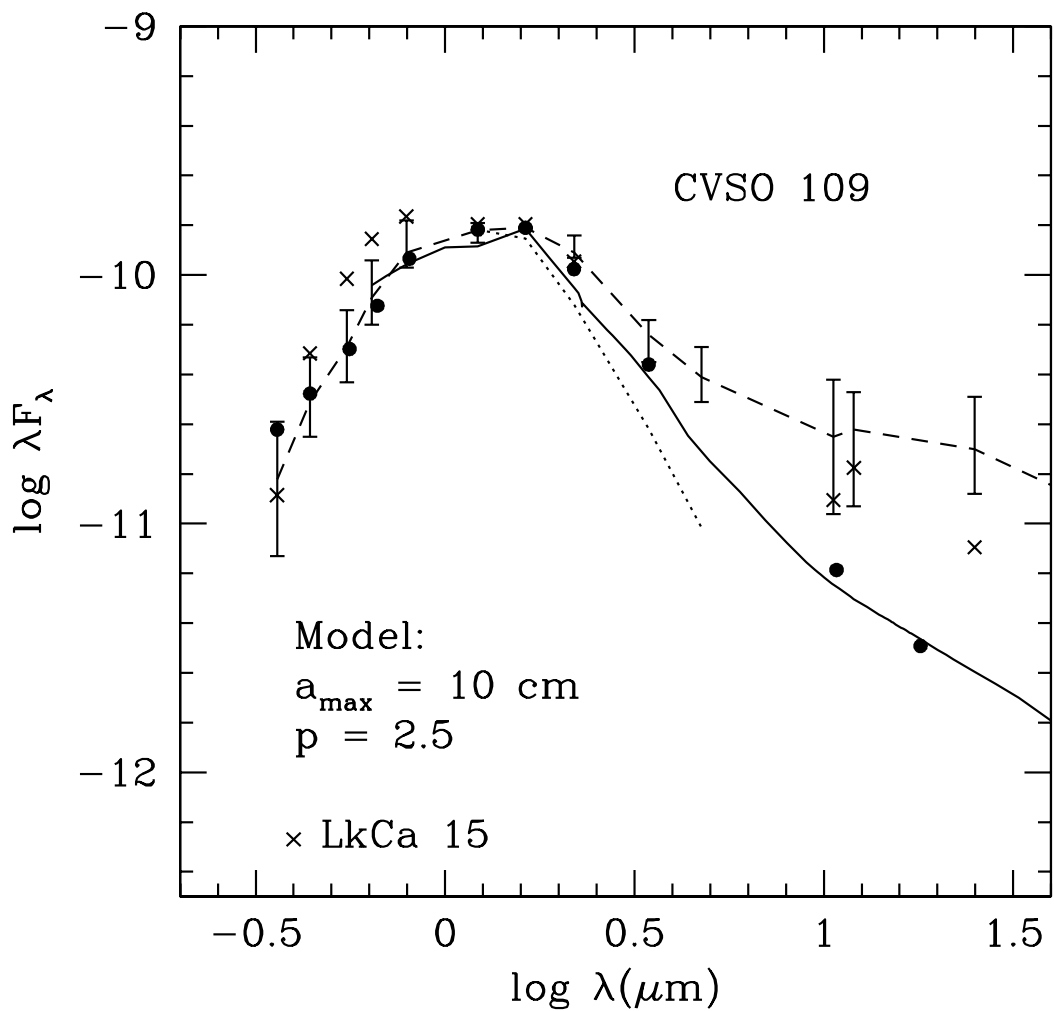}
\caption{Spectral energy distribution of
CVSO 109 (solid filled circles) compared
to the median of Taurus (dashed lines),
the photospheric fluxes (dotted lines),
and the
SED of LkCa 15 (x's). The solid
line shows the SED of an irradiated accretion
disk model with similar stellar and accretion
parameters as CVSO 109, and with
a dust size distribution characterized by
$a_{max} = $ 10 cm and exponent $p$ = 2.5 (see
text).
}
\label{fig_est109}
\end{figure}

A number of claims of grain growth in disks have
been made over the years, starting with the
interpretation of slopes of the fluxes in the
millimeter, which are expected to become
flatter as the grain size increases (Beckwith \& Sargent 1991).
Assuming a dust mixture and a
dust size distribution of the form
$n(a) \propto a^{-p}$ between minimum
and maximum sizes $a_{min}$ and $a_{max}$,
D01 find that a much better fit to the median SED of Taurus can
be achieved with
dust particle growth  to a maximum size $a_{max}$
of 1 mm ($p$=3.5) than with 
interstellar grains, with $a_{max} = 0.25 \mu$m.
Grain growth and settling has been claimed
to explain the SEDs of some CTTS in Taurus. 
One well studied case is that of the
CTTS LkCa 15,
which has a SED much flatter than the Taurus median 
in the IR and high
emission at submillimeter and millimeter wavelengths
(Qi et al. 2003); the interpretation of this
SED requires significant grain growth and settling
(Chiang et al. 2001;
D'Alessio et al. 2003, 2004; Bergin et al. 2004).
In Figures
\ref{figviuv}, \ref{figviub},
\ref{fighkub}, \ref{figklub}, 
and \ref{figknkl} 
we have marked the location of LkCa 15;
disks in Ori OB1 have similar colors
to those of LkCa 15 in all bands, suggesting
that significant dust evolution has also
taken place in these older disks.
In Figure \ref{fig_est109} we compare
the SED of CVSO 109, the
only object for which we have
detections up to 18$\mu$m, to
fluxes for LkCa 15 from KH95
and to the Taurus median,
both scaled at the $H$ of CVSO 109. Note
that LkCa 15 is a K5 star (KH95), so
it is brighter in the optical
than CVSO 109, which is an M0 (Table \ref{tab_prop}). 
The SEDs of LkCa 15 and CVSO 109 are significantly lower than the
Taurus median, although the flux deficit is
much larger in CVSO 109.

To try to quantify the degree of dust evolution
in CVSO 109, we have calculated models 
with uniformly distributed gas and dust
following the procedures of D01.
We have used stellar and accretion parameters from
Table \ref{tab_prop}, and
have varied the parameters $a_{max}$
and $p$ which characterize
the dust size distribution. Figure \ref{fig_est109} shows
our best model. To fit the observed
fluxes, we require $a_{max}$ = 10 cm; 
this is higher than the maximum sizes required to explain the median SED
in Taurus, $a_{max} \sim$  1 mm (D01),
or the SED of the $\sim$ 10 Myr old star TW Hya,
$a_{max} \sim$  1 cm (Calvet et al. 2002; Wilner et al. 2004).
Moreover, we require $p = 2.5$, a value 
consistent with significant dust coagulation
(Miyake \& Nakagawa 1995); 
modeling of CTTS disks can generally be
carried out successfully with the
assumption 
of $p = $ 3.5, a value characteristic of the
interstellar medium (cf. D01, Chiang et al. 2001).

Our modeling is only indicative, in that we
have assumed that the dust is uniformly
distributed with the gas. Even with this
assumption,
our comparison implies that the disk of CVSO 109 is
essentially a flat disk, with little
if any flaring. Thus, our modeling
indicates that the dust in the disk of CVSO 109
has strongly settled towards the midplane,
as expected from theories of 
dust evolution in protoplanetary disks
(Weidenschilling 1997).
Models including settling in addition to observations
at longer wavelengths are required to
actually constrain the size distribution
of the solids in the midplane.

\begin{figure}
\plotone{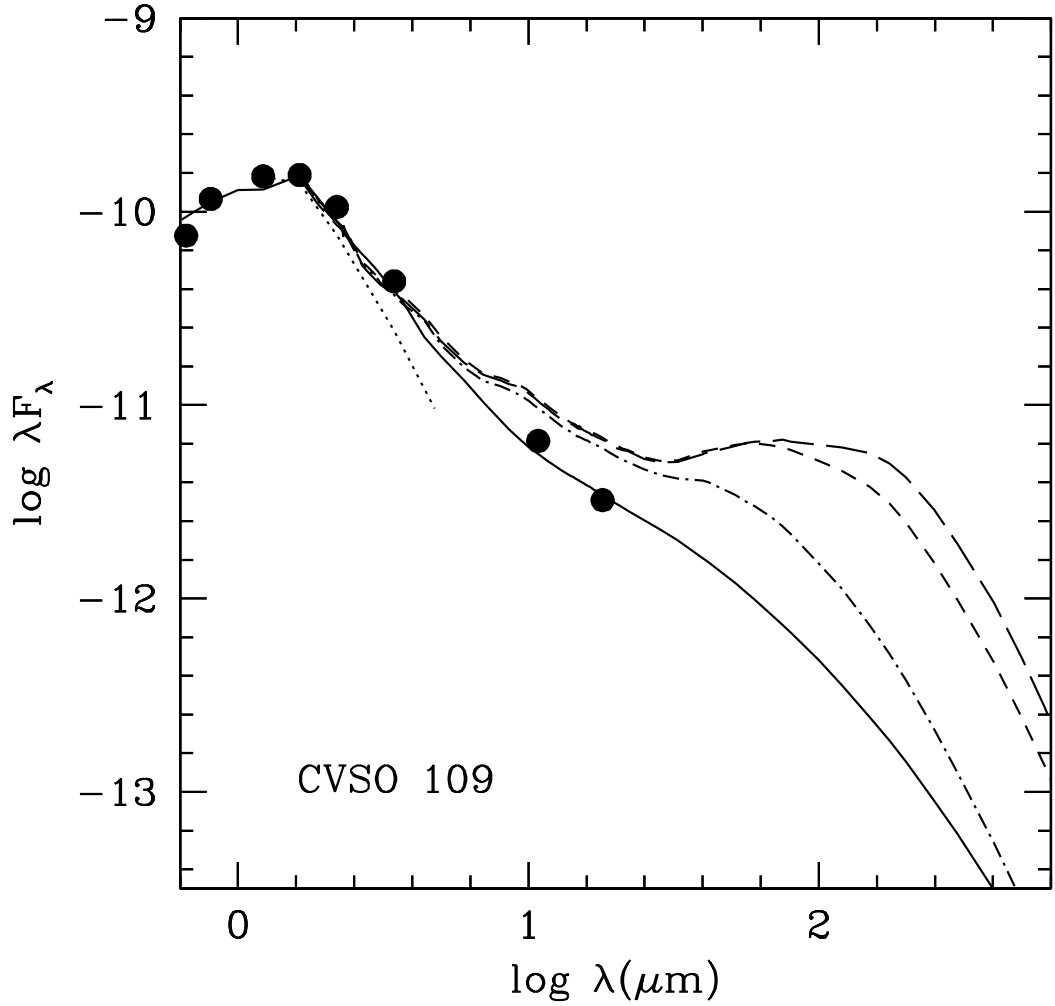}
\caption{Effect of disk radius on SED. Disk models
for $a_{max} $ = 1mm and $p$ = 3.5,  and disk radius
100 AU (long-dashed line), 50 AU (short-dashed line), and 10 AU (dot-dashed line).
Model for $a_{max} $ = 10 cm and $p$ = 2.5 in solid line
and observations of CVSO 109 in filled circles.
The photosphere is indicated by the dotted line.
}
\label{figsmalldisk}
\end{figure}

A possible alternative to consider is that disks
in Ori OB1
are smaller than their Taurus counterparts
because their outer parts have been photoevaporated
by the external ionizing fields of the high
mass stars in the associations. A smaller disk would
have less emission at long wavelengths since
the low temperature regions would be removed.
Models for the viscous evolution
of disks subject to the effects of photoevaporating
radiation fields from the accretion flows and
external radiation fields
have been calculated by 
Clarke, Gendrin, \& Sotomayor (2001) and
Matsuyama, Johnston \& Hartmann (2003).
Since the distance
from the ionizing stars are
of the order of a few pc in OB associations, FUV
external photons are expected to dominate over EUV photons
in the photoevaporation of disks
(Johnstone et al. 1998). 
For this case, Matsuyama et al. (2003)
find that
that the outer portions of the disk could be removed
on time scales which range from 1 to several Myr,
depending on the viscosity parameter. 
After significant photoevaporation, 
the disk edge is fixed by the FUV gravitational
radius, which for parameters as those
of the Ori OB1b stellar sample is  $\sim$ 30 AU.
To determine the effects
of a reduction of disk radius on the SED,
and in particular to see if
a smaller disk radius can explain the low
mid-infrared fluxes without extreme grain growth,
we have calculated disk models with $a_{max}$ = 1 mm
and $p$ = 3.5, characteristic of the Taurus
population, for disk radii 100 AU, 50 AU, and 10 AU.
These SEDs are shown in 
Figure \ref{figsmalldisk}.
Models are calculated at a high inclination
to the line of sight ($72^{\circ}$) to maximize
the contrast between the star and the disk.
Even small disks produce too much
long wavelength flux compared to the
the observations, if $a_{max}$ = 1 mm
as in Taurus. This implies
again that the disk has to be
essentially flat, which require that grains 
grow to larger sizes and/or have settled.

\section{Summary and conclusions}
\label{sec_concl}

We have detected clear signatures of disk
evolution in samples of accreting stars in 
the Ori OB1a and 1b associations, with
ages 7 - 10 Myr and 3 - 5 Myr,
respectively. 
The mass accretion rates of these stars
are consistent with viscous evolution of
accretion disks. In addition, we find a
significant overall decrease of infrared
emission with age. Comparison with disk
models indicates 
that grain growth and settling, processes expected
from dust evolution,  are
responsible for this decrease.
Future observations of larger
samples of Ori OB1 stars with SPITZER
will provide more detail
on the nature of the dust evolution process.

Acknowledgments. 
We thank the referee for many useful suggestions.
This work was supported by
NSF grant AST-9987367 and NASA grant NAG5-10545.
PD acknowledges grants from PAPIIT, DGAPA (UNAM) and
CONACyT, M\'exico.
This research makes use of data products 
from the Two Micron All Sky Survey, which is a joint project 
of the University of Massachusetts and the Infrared 
Processing and Analysis Center/California Institute of 
Technology, funded by the National Aeronautics and Space Administration 
and the National Science Foundation.
This paper is based on observations obtained with the mid-infrared camera 
OSCIR, developed by the University of Florida with support from the National 
Aeronautics and Space Administration, and operated jointly by Gemini and the 
University of Florida Infrared Astrophysics Group.
This research has made use of the SIMBAD database,
operated at CDS, Strasbourg, France.

{}

\clearpage
 
\begin{deluxetable}{lccccccccc}
\tablewidth{0pt} 
\tablecaption{OPTICAL AND NEAR INFRARED PHOTOMETRY \label{tab_photometry}}
\label{fotometria} 
\tablehead{
\colhead{CVSO} & \colhead{V}  & \colhead{B-V} & \colhead{U-V} &  \colhead{V-R} & \colhead{V-I} &  \colhead{J} & \colhead{J-H}&\colhead{H-K$_s$} &\colhead{L}  \\
}
\startdata
   1&  17.12&  1.36& -0.36&  1.26&  2.81& 12.82&  0.68&  0.25& \nodata  \\
   3&  16.31&  1.47&  0.99&  1.06&  2.25& 12.69&  0.67&  0.26& \nodata  \\
   5&  16.09&  1.48&  0.81&  1.08&  2.23& 12.50&  0.68&  0.22& \nodata  \\
  10&  15.41&  1.42&  1.27&  0.93&  1.91& 12.31&  0.68&  0.24& \nodata  \\
  11&  15.74&  1.53&  0.96&  1.08&  2.33& 12.10&  0.66&  0.23& \nodata  \\
  15&  15.95&  1.44&  0.46&  0.90&  2.12& 12.57&  0.74&  0.19& \nodata  \\
  17&  16.16&  1.51& \nodata &  1.07&  2.34& 12.52&  0.72&  0.20& \nodata  \\
  19&  15.89&  1.49&  1.54&  1.14&  2.51& 11.92&  0.69&  0.21& \nodata  \\
  22&  16.33&  1.51&  0.86&  1.06&  2.29& 12.70&  0.66&  0.22& \nodata  \\
  24&  16.31&  1.49& \nodata &  1.06&  2.25& 12.79&  0.72&  0.14& \nodata  \\
  25&  15.46&  1.42& \nodata &  0.95&  1.91& 12.42&  0.69&  0.15& \nodata  \\
  29&  16.25&  1.51& \nodata &  1.14&  2.53& 12.38&  0.67&  0.24& \nodata  \\
  34&  15.15&  1.36&  1.16&  0.96&  1.97& 12.28&  0.99&  0.23& \nodata  \\
  34&  15.15&  1.36&  1.16&  0.96&  1.97& 12.72&  1.34&  0.23& \nodata  \\
  35&  14.55&  1.41&  0.73&  0.99&  1.95& 11.46&  0.70&  0.41& \nodata  \\
    &  14.84&  1.30&  0.76&  0.92&  1.93& 11.46&  0.70&  0.41& \nodata  \\
  36&  16.07&  1.45& \nodata &  1.09&  2.35& 12.38&  0.69&  0.21& \nodata  \\
  38&  16.24&  1.57&  0.99&  1.07&  2.27& 12.79&  0.73&  0.23& \nodata  \\
  38&  16.37&  1.26&  1.12&  1.07&  2.46& 12.79&  0.73&  0.23& \nodata  \\
  39&  15.98&  1.41&  0.90&  1.07&  2.45& 12.30&  0.68&  0.20& \nodata  \\
  39&  16.11&  1.22&  0.66&  1.10&  2.46& 12.30&  0.68&  0.20& \nodata  \\
  40&  15.23&  1.35&  1.07&  0.90&  1.81& 12.36&  0.72&  0.26& \nodata  \\
  41&  15.23&  0.97&  0.28&  0.73&  1.54& 12.41&  0.70&  0.33& \nodata  \\
  42&  15.96&  1.43&  1.12&  1.03&  2.15& 12.68&  0.68&  0.17& \nodata  \\
  43&  16.27&  1.42&  0.98&  1.09&  2.46& 12.50&  0.74&  0.16& \nodata  \\
  44&  14.74&  1.21&  0.96&  0.73&  1.43& 12.30&  0.66&  0.08& \nodata  \\
  46&  15.15&  1.42& \nodata &  0.94&  1.93& 12.02&  0.69&  0.21& \nodata  \\
  46&  15.19&  1.46&  0.91&  0.94&  1.92& 12.02&  0.69&  0.21& \nodata  \\
  47&  15.09&  1.38&  1.12&  0.93&  1.86& 11.28&  0.84&  0.50&  9.36 \\
  48&  14.07&  1.16&  0.70&  0.78&  1.44& 11.50&  0.68&  0.22& 10.45 \\
  54&  15.80&  1.31&  1.18&  0.91&  1.95& 12.86&  0.70&  0.14& \nodata  \\
  55&  15.60&  1.43&  1.16&  1.06&  2.27& 11.93&  0.69&  0.24& \nodata  \\
  55&  15.69&  1.54&  0.99&  1.01&  2.26& 11.93&  0.69&  0.24& \nodata  \\
  57&  15.67& \nodata & \nodata &  0.91&  1.91& 12.65&  0.72&  0.19& \nodata  \\
  58&  14.93&  0.89& -0.30&  0.84&  1.65& 12.05&  0.78&  0.56&  9.97 \\
  72&  17.98&  1.64&  0.87&  0.94&  2.22& 14.49&  0.71&  0.21& \nodata  \\
  75&  16.72&  0.92&  0.41&  0.44&  1.04& 13.00&  0.76&  0.51& \nodata  \\
  77&  16.37&  1.42& \nodata &  1.07&  2.26& 12.81&  0.72&  0.23& \nodata  \\
  81&  16.22&  1.45&  0.87&  1.17&  2.54& 12.24&  0.66&  0.23& \nodata  \\
  90&  14.61&  0.38& -0.97&  0.65&  1.40& 12.13&  0.91&  0.67&  9.73 \\
 104&  14.22&  0.79& -0.08&  0.66&  1.44& 11.78&  0.75&  0.58& \nodata  \\
 107&  14.78&  0.84& -0.38&  0.81&  1.73& 11.55&  0.81&  0.45&  9.85 \\
 109&  13.97&  0.88& -0.35&  0.80&  1.69& 11.01&  0.80&  0.42&  9.37 \\
 121&  14.24&  1.18&  0.23&  0.87&  1.69& 11.70&  0.92&  0.64& \nodata  \\
 124&  15.91&  1.50&  0.89&  1.11&  2.41& 12.07&  0.70&  0.25& \nodata  \\
 126&  15.82&  1.44&  0.99&  0.92&  1.85& 12.85&  0.71&  0.17& \nodata  \\
 133&  14.88&  1.15&  1.01&  0.76&  1.59& 12.19&  0.61&  0.15& \nodata  \\
 143&  15.25&  1.34&  0.41&  1.03&  2.19& 11.79&  0.78&  0.21& \nodata  \\
 146&  13.92&  1.19&  0.14&  0.84&  1.63& 11.30&  0.86&  0.42&  9.65 \\
    &  14.01& \nodata & \nodata & \nodata &  1.52& 11.30&  0.86&  0.42&  9.65 \\
 152&  13.98&  1.12&  0.81&  0.71&  1.40& 11.58&  0.82&  0.50& \nodata  \\
 153&  16.00&  1.43& -0.64&  1.07&  2.29& 12.44&  0.72&  0.23& \nodata  \\
 155&  14.35&  1.33&  0.63&  0.92&  1.85& 11.48&  0.66&  0.33& \nodata  \\
 157&  17.87&  0.70& -0.98&  1.14&  2.09& 14.01&  0.59&  0.35& \nodata  \\
 164&  16.71&  1.55&  0.30&  1.31&  3.03& 11.80&  0.69&  0.32& \nodata  \\
 165&  13.68&  1.12&  0.45&  0.76&  1.48& 11.07&  0.83&  0.41&  9.24 \\
 168&  17.33&  1.24& -0.25&  1.27&  2.82& 12.78&  0.60&  0.28& \nodata  \\
 171&  14.84&  1.46&  0.64&  0.84&  1.69& 11.96&  0.70&  0.30& \nodata  \\
 176&  15.65&  1.21& -0.35&  1.14&  2.46& 11.71&  0.91&  0.50&  9.84 \\
 177&  15.62&  0.71& \nodata &  0.96&  1.92& 12.30&  0.77&  0.33& 10.98 \\
 178&  15.66&  1.48&  0.64&  1.11&  2.28& 11.47&  0.77&  0.34& \nodata  \\
 181&  15.63&  1.22& \nodata &  0.87&  2.00& 12.30&  0.74&  0.19& \nodata  \\
 184&  15.34&  1.86& \nodata &  1.41&  2.75& 10.40&  1.27&  0.91& \nodata  \\
 185&  15.28&  1.61& \nodata &  1.08&  2.18& 10.88&  0.88&  0.49&  8.77 \\
 190&  14.04&  1.10&  0.28&  0.83&  1.72& 11.04&  1.08&  0.77&  8.49 \\
    &  14.76& \nodata & \nodata & \nodata &  1.89& 11.04&  1.08&  0.77&  8.49 \\
 192&  15.99&  1.26& -0.14&  1.17&  2.37& 11.56&  1.01&  0.55& 99.00 \\
 193&  15.15&  1.29&  0.04&  1.15&  2.29& 11.34&  1.22&  0.66& 99.00 \\
\enddata
\end{deluxetable}

\begin{deluxetable}{lcccccc}
\tablewidth{0pt}
\tablecaption{OSCIR PHOTOMETRY \label{tab_oscir}}
\label{mid-irfotometria}
\tablehead{
\colhead{CVSO} &\colhead{$N$}  & \colhead{err$N$} & \colhead{[18$\mu$m]} &  \colhead{err[18$\mu$m]} & \colhead{log $\lambda F_{\lambda}(N)$} &  \colhead{log $\lambda F_{\lambda}(18\mu$m)} \\
\colhead{} &\colhead{}  & \colhead{} & \colhead{} &  \colhead{} & \colhead{log (erg sec$^{-1}$ cm$^{-2}$)} &  \colhead{log (erg sec$^{-1}$ cm$^{-2}$)}
}
\startdata
47 & 9.418 & 0.081 &  $>$ 7.627&  0.383 &   -11.779 $\pm$ 0.03 &  $<$ -11.654   \\
58 & 8.448 & 0.388 &  $>$ 7.517&  0.444 &   -11.391$\pm$  0.15 &  $<$ -11.576  \\
90 & 8.294 & 0.282 &  $>$ 9.533&  2.773 &   -11.330$\pm$  0.12 &  $<$ -11.792  \\
107&  8.109&  0.276&  $>$  9.092&  1.866&    -11.256$\pm$  0.12&   $<$ -11.764  \\
109&  7.935&  0.120&   6.437&  0.165&    -11.186$\pm$  0.05&    -11.492$\pm$  0.1  \\
165&  6.921&  0.093&  $>$  6.983&  0.279&    -10.781$\pm$  0.05&   $<$ -11.710 \\
\enddata
\end{deluxetable}

\begin{deluxetable}{lccccccccccc}
 \tabletypesize{\scriptsize}
\tablewidth{0pt}
\tablecaption{Stellar and Accretion Properties \label{tab_prop}}
\label{prop}
\tablehead{
\colhead{CVSO} & \colhead{SpT}  & \colhead{$T_{eff}$} & \colhead{$L$} &  \colhead{$R$} & \colhead{$M$} &  \colhead{$A_V$} & \colhead{Age}&\colhead{$L_{acc}$} &\colhead{$\mdot$} &\colhead{Assoc} &\colhead{Type}  \\
\colhead{} & \colhead{}  & \colhead{K} & \colhead{$\lsun$} &  \colhead{$\rsun$} & \colhead{$\msun$} &  \colhead{} & \colhead{Myr}&\colhead{$\lsun$} &\colhead{$10^{-8} \, \msunyr$} &\colhead{} &\colhead{} \\
}
\startdata
   1&  M3&3470&  0.18&  1.17&  0.33&  0.81&  4.02&  0.01&  0.09&    1a&  CTTS \\
   3&  M3&3470&  0.16&  1.11&  0.33&  0.00&  4.68&  0.00&  0.00&    1a&  WTTS \\
   5&  M2&3580&  0.21&  1.20&  0.38&  0.22&  4.12&  0.00&  0.01&    1a&  WTTS \\
  10&  M0&3850&  0.27&  1.17&  0.58&  0.27&  7.20&  0.00&  0.00&    1a&  WTTS \\
  11&  M2&3580&  0.33&  1.49&  0.39&  0.46&  2.51&  0.00&  0.02&    1a&  WTTS \\
  15&  M1&3720&  0.21&  1.11&  0.47&  0.39&  6.42&  0.00&  0.03&    1a&  WTTS \\
  17&  M3&3470&  0.19&  1.20&  0.33&  0.00&  3.63&  0.00&  0.06&    1a&  WTTS \\
  19&  M3&3470&  0.34&  1.60&  0.34&  0.10&  2.23& \nodata   & \nodata   &    1a&  WTTS \\
  22&  M2&3580&  0.18&  1.12&  0.38&  0.36&  5.18&  0.00&  0.01&    1a&  WTTS \\
  24&  M2&3580&  0.17&  1.06&  0.38&  0.27&  5.97&  0.00&  0.01&    1a&  WTTS \\
  25&  M0&3850&  0.24&  1.11&  0.58&  0.27&  8.09& \nodata   & \nodata   &    1a&  WTTS \\
  29&  M2&3580&  0.29&  1.40&  0.39&  0.94&  2.73& \nodata   & \nodata   &    1a&  WTTS \\
  34&  M0&3850&  0.29&  1.21&  0.58&  0.41&  6.61&  0.00&  0.01&    1a&  WTTS \\
  34&  M0&3850&  0.19&  0.98&  0.59&  0.41& 10.55&  0.00&  0.01&    1a&  WTTS \\
  35&  K7&4060&  0.72&  1.72&  0.77&  0.85&  2.88&  0.02&  0.17&    1a&  CTTS \\
    &  K7&4060&  0.71&  1.70&  0.77&  0.80&  2.90&  0.02&  0.13&    1a&  CTTS \\
  36&  M3&3470&  0.21&  1.28&  0.34&  0.00&  2.95&  0.03&  0.36&    1a&  WTTS \\
  38&  M2&3580&  0.17&  1.06&  0.38&  0.31&  5.88&  0.00&  0.00&    1a&  WTTS \\
  38&  M2&3580&  0.19&  1.13&  0.38&  0.77&  4.98&  0.00&  0.02&    1a&  WTTS \\
  39&  M2&3580&  0.30&  1.41&  0.39&  0.74&  2.70&  0.00&  0.04&    1a&  WTTS \\
  39&  M2&3580&  0.30&  1.42&  0.39&  0.77&  2.68&  0.01&  0.08&    1a&  WTTS \\
  40&  M0&3850&  0.24&  1.10&  0.58&  0.02&  8.18&  0.00&  0.00&    1a&  CTTS \\
  41&  K2&4900&  0.37&  0.84&  0.89&  1.30& 88.18&  0.06&  0.19&    1a&  CTTS \\
  42&  M2&3580&  0.17&  1.08&  0.38&  0.02&  5.71&  0.00&  0.00&    1a&  WTTS \\
  43&  M2&3580&  0.25&  1.29&  0.39&  0.77&  3.03&  0.00&  0.02&    1a&  WTTS \\
  44&  K6&4205&  0.27&  0.98&  0.82&  0.00& 22.97&  0.00&  0.01&    1a&  WTTS \\
  46&  M0&3850&  0.36&  1.34&  0.58&  0.32&  4.74&  0.00&  0.03&    1a&  WTTS \\
  46&  M0&3850&  0.35&  1.34&  0.58&  0.29&  4.80&  0.00&  0.02&    1a&  WTTS \\
  47&  K5&4350&  0.97&  1.73&  1.10&  1.21&  4.96&  0.01&  0.04&    1a&  CTTS \\
  48&  K4&4590&  1.25&  1.77&  1.28&  0.71&  6.60&  0.05&  0.22&    1b&  CTTS \\
  54&  M0&3850&  0.17&  0.92&  0.58&  0.36& 14.57&  0.00&  0.01&    1a&  WTTS \\
  55&  M2&3580&  0.37&  1.58&  0.39&  0.31&  2.30&  0.00&  0.01&    1a&  WTTS \\
  55&  M2&3580&  0.37&  1.58&  0.39&  0.29&  2.31&  0.00&  0.01&    1a&  WTTS \\
  57&  M0&3850&  0.20&  1.00&  0.59&  0.27&  9.95&  0.00&  0.00&    1a&  CTTS \\
  58&  K7&4060&  0.61&  1.58&  0.78&  0.12&  4.02&  0.07&  0.45&    1b&  CTTS \\
  72&  M6&3050&  0.04&  0.74&  0.11&  0.00&  6.20&  0.00&  0.01&    1b&  CTTS \\
  75&  M1&3720&  0.23&  1.16&  0.47&  0.00&  5.82&  0.00&  0.04&    1b&  CTTS \\
  77&  K6&4205&  0.28&  0.99&  0.82&  1.88& 21.91& \nodata   & \nodata   &    1a&  WTTS \\
  81&  M3&3470&  0.45&  1.86&  0.34&  0.17&  1.76&  0.00&  0.02&    1b&  WTTS \\
  90&  K7&4060&  0.55&  1.50&  0.79&  0.00&  5.04&  0.29&  1.77&    1b&  CTTS \\
 104&  K7&4060&  0.76&  1.76&  0.77&  0.00&  2.76&  0.10&  0.75&    1b&  CTTS \\
 107&  K7&4060&  1.02&  2.04&  0.77&  0.32&  2.12&  0.13&  1.09&    1b&  CTTS \\
 109&  M0&3850&  1.48&  2.73&  0.56&  0.00&  0.80&  0.16&  2.52&    1b&  CTTS \\
 121&  K3&4730&  1.31&  1.70&  1.32&  1.49&  8.06&  0.40&  1.67&    1b&  CTTS \\
 124&  M3&3470&  0.51&  1.97&  0.34&  0.00&  1.58&  0.00&  0.02&    1b&  WTTS \\
 126&  M0&3850&  0.16&  0.89&  0.58&  0.12& 16.16&  0.00&  0.00&    1a&  WTTS \\
 133&  K6&4205&  0.32&  1.06&  0.84&  0.27& 18.14&  0.00&  0.01&    1a&  WTTS \\
 143&  M1&3720&  0.82&  2.18&  0.47&  0.55&  1.16&  0.02&  0.37&    1b&  CTTS \\
 146&  K6&4205&  1.32&  2.17&  0.91&  0.37&  2.01&  0.11&  0.81&    1b&  CTTS \\
    &  K6&4205&  1.23&  2.09&  0.92&  0.10&  2.19&  0.12&  0.88&    1b&  CTTS \\
 152&  K3&4730&  1.21&  1.64&  1.30&  0.78&  8.65&  0.03&  0.14&    1b&  CTTS \\
 153&  M2&3580&  0.23&  1.26&  0.39&  0.36&  3.43&  0.02&  0.16&    1a&  WTTS \\
 155&  K6&4205&  1.30&  2.14&  0.91&  0.90&  2.06&  0.07&  0.53&    1b&  CTTS \\
 157&  C      &\nodata& \nodata&\nodata&\nodata& \nodata& \nodata& \nodata& \nodata & 1b&  CTTS \\
 164&  M3&3470&  1.06&  2.85&  0.35&  1.82&  0.45&  0.05&  1.19&    1b&  CTTS \\
 165&  K6&4205&  1.48&  2.29&  0.91&  0.00&  1.72&  0.05&  0.37&    1b&  CTTS \\
 168&  M4&3370&  0.25&  1.46&  0.29&  0.00&  2.49&  0.00&  0.05&    1b&  CTTS \\
 171&  K5&4350&  0.82&  1.60&  1.08&  0.81&  6.41&  0.02&  0.10&    1b&  CTTS \\
 176&  M3&3470&  0.71&  2.32&  0.35&  0.00&  1.06&  0.02&  0.43&    1b&  CTTS \\
 177&  K6&4205&  0.64&  1.50&  0.94&  1.07&  6.21&  0.01&  0.04&    1b&  CTTS \\
 178&  M1&3720&  1.17&  2.60&  0.47&  0.77&  0.81&  0.01&  0.25&    1b&  CTTS \\
 181&  M0&3850&  0.51&  1.61&  0.57&  0.48&  2.63&  0.02&  0.14&    1b&  CTTS \\
 184&  K1&5080&  9.67&  4.01&  2.35&  4.35&  2.03& \nodata   & \nodata   &    1b&  CTTS \\
 185&  K7&4060&  2.54&  3.22&  0.81&  1.40&  0.79& \nodata   & \nodata   &    1b&  CTTS \\
 190&  K6&4205&  1.78&  2.52&  0.90&  0.58&  1.26&  0.13&  1.17&    1b&  CTTS \\
    &  K6&4205&  1.99&  2.66&  0.90&  0.99&  1.00&  0.14&  1.38&    1b&  CTTS \\
 192&  K6&4205&  1.68&  2.44&  0.90&  2.14&  1.40&  0.36&  3.08&    1c&  CTTS \\
 193&  K6&4205&  1.96&  2.63&  0.90&  1.95&  1.02&  0.48&  4.49&    1c&  CTTS \\
\enddata
\end{deluxetable}

\clearpage

\clearpage

\clearpage

\clearpage

\clearpage

\clearpage

\clearpage

\clearpage

\clearpage

\clearpage

\clearpage


\begin{thebibliography}{}

\bibitem[Alibert, Mordasini, \& Benz(2004)]{2004A&A...417L..25A} Alibert, 
Y., Mordasini, C., \& Benz, W.\ 2004, \aap, 417, L25 

\bibitem[Bally, Stark, Wilson, \& Langer(1987)]{1987ApJ...312L..45B} Bally, 
J., Stark, A.~A., Wilson, R.~W., \& Langer, W.~D.\ 1987, \apjl, 312, L45 

\bibitem[Barden \& Armandroff(1995)]{bard95} Barden, S. \& Armandroff, T., 1995: Hydra/WIYN users manual.

\bibitem[Beckwith \& Sargent(1991)]{1991ApJ...381..250B} Beckwith, 
S.~V.~W.~\& Sargent, A.~I.\ 1991, \apj, 381, 250 

Bergin, E. and 12 coauthors, 2004, ApJL, submitted.


\bibitem[Blaauw(1964)]{1964ARA&A...2..213B} Blaauw, A.\ 1964, \araa, 2, 213

\bibitem[Brice\~{n}o et al.(2001)]{bric01} Brice\~{n}o, C. and 10 coauthors, 2001, Sci, 291, 93

\bibitem[Brice\~{n}o et al.(2004)]{bric04} Brice\~{n}o, C.,
Calvet, N., Hernandez, J., Vivas, A.K., Hartmann, L., Downes, J.J., 
\& Berlin, P. 2004, AJ, submitted (Paper I)

\bibitem[Brown et al.(1994)]{bgz94}  Brown, A.G.A.,  de Geus, E.J., \& de Zeeuw, P.T. 1994, AA, 289, 101

\bibitem[Calvet \& Gullbring(1998)]{calv98} Calvet, N. \& Gullbring, E., 1998, \apj, 509,802

\bibitem[Calvet et al.(2002)]{calv02} Calvet, N., D'Alessio, P., Hartmann, L., Wilner, D., Walsh, A. \& 
Sitko, M., 2002, \apj, 568, 1008

\bibitem[Cardelli et al.(1989)]{carde89} Cardelli,J.A.,  Clayton, G. C., Mathis, J.S. 1989,  \apj , 345, 24
5

\bibitem[Carpenter(2001)]{2001AJ....121.2851C} Carpenter, J.~M.\ 2001, \aj, 
121, 2851

\bibitem[Chiang et al.(2001)]{2001ApJ...547.1077C} Chiang, E.~I., Joung, 
M.~K., Creech-Eakman, M.~J., Qi, C., Kessler, J.~E., Blake, G.~A., \& van 
Dishoeck, E.~F.\ 2001, \apj, 547, 1077 

\bibitem[Clarke, Gendrin, \& Sotomayor(2001)]{2001MNRAS.328..485C} Clarke, 
C.~J., Gendrin, A., \& Sotomayor, M.\ 2001, \mnras, 328, 485 

\bibitem[Cohen et al. (1999)]{cwc99}Cohen, M., Walker, R. G., Carter, B.,
    Hammersley, P., Kidger, M., \& Noguchi, K. 1999, \aj, 117, 1864

\bibitem[Colavita et al.(2003)]{2003ApJ...592L..83C} Colavita, M., et al.\ 
2003, \apjl, 592, L83 

D'Alessio, P. and 13 coauthors, 2004, ApJ, submitted

\bibitem[D'Alessio, Calvet, \& Hartmann(2001)]{2001ApJ...553..321D} 
D'Alessio, P., Calvet, N., \& Hartmann, L.\ 2001, \apj, 553, 321, D01 

\bibitem[D'Alessio et al.(1999)]{1999ApJ...527..893D} D'Alessio, P., 
Calvet, N., Hartmann, L., Lizano, S., \& Cant{\' o}, J.\ 1999, \apj, 527, 
893 

\bibitem[D'Alessio, Canto, Calvet, \& Lizano(1998)]{1998ApJ...500..411D} 
D'Alessio, P., Canto, J., Calvet, N., \& Lizano, S.\ 1998, \apj, 500, 411 

\bibitem[D'Antona \& Mazzitelli(1994)]{1994ApJS...90..467D} D'Antona, F.~\& 
Mazzitelli, I.\ 1994, \apjs, 90, 467 

\bibitem[Dolan \& Mathieu(2002)]{2002AJ....123..387D} Dolan, C.~J.~\& 
Mathieu, R.~D.\ 2002, \aj, 123, 387 

\bibitem[Dolan \& Mathieu(2001)]{2001AJ....121.2124D} Dolan, C.~J.~\& 
Mathieu, R.~D.\ 2001, \aj, 121, 2124 

\bibitem[Dolan \& Mathieu(1999)]{1999AJ....118.2409D} Dolan, C.~J.~\& 
Mathieu, R.~D.\ 1999, \aj, 118, 2409 

\bibitem[Dullemond, Dominik, \& Natta(2001)]{2001ApJ...560..957D} 
Dullemond, C.~P., Dominik, C., \& Natta, A.\ 2001, \apj, 560, 957 

\bibitem[Elias et al. (1982)]{efm82} Elias, J.H., Frogel, J.A., Matthews, K.
        \& Neugebauer, G. 1982, \aj, 87, 1893

\bibitem[de Zeeuw et al.(1999)]{1999AJ....117..354D} de Zeeuw, P.~T., 
Hoogerwerf, R., de Bruijne, J.~H.~J., Brown, A.~G.~A., \& Blaauw, A.\ 1999, 
\aj, 117, 354 

\bibitem[Fabricant et al.(1998)]{fab98} Fabricant, D.,  Cheimets, P., Caldwell, N. \& Geary, J., 1998, \pasp, 110, 79

\bibitem[Gullbring et al.(1998)]{gull98} Gullbring, E., Hartmann, L., Brice\~{n}o, C., Calvet, N., 1998, ApJ 492, 323

\bibitem[Gullbring, Calvet, Muzerolle, \& 
Hartmann(2000)]{2000ApJ...544..927G} Gullbring, E., Calvet, N., Muzerolle, 
J., \& Hartmann, L.\ 2000, \apj, 544, 927 

\bibitem[Hartigan, Edwards, \& Ghandour(1995)]{1995ApJ...452..736H} 
Hartigan, P., Edwards, S., \& Ghandour, L.\ 1995, \apj, 452, 736 

\bibitem[Hartigan et al.(1991)]{1991ApJ...382..617H} Hartigan, P., Kenyon, 
S.~J., Hartmann, L., Strom, S.~E., Edwards, S., Welty, A.~D., \& Stauffer, 
J.\ 1991, \apj, 382, 617 

\bibitem[Hartmann(1998)]{hart98} Hartmann, L.: Accretion Processes in Star Formation, Cambridge University
Press, 1998.

\bibitem[Hartmann et al.(1998)]{hartal98} Hartmann, L., Calvet, N., Gullbring, E. \& D'Alessio, P, 1998, \aj, 495, 385

\bibitem[Hartmann(2003)]{hart03} Hartmann, L., 2003, \apj, 585, 398

\bibitem[Hillenbrand, Strom, Vrba, \& Keene(1992)]{1992ApJ...397..613H} 
Hillenbrand, L.~A., Strom, S.~E., Vrba, F.~J., \& Keene, J.\ 1992, \apj, 
397, 613

\bibitem[Hillenbrand(1997)]{hill97} Hillenbrand, L.A., 1997, \aj, 113, 5

\bibitem[Johnstone, Hollenbach, \& Bally(1998)]{1998ApJ...499..758J} 
Johnstone, D., Hollenbach, D., \& Bally, J.\ 1998, \apj, 499, 758

\bibitem[Kenyon \& Hartmann(1987)]{1987ApJ...323..714K} Kenyon, S.~J.~\& 
Hartmann, L.\ 1987, \apj, 323, 714 

\bibitem[Kenyon \& Hartmann(1995)]{ken95}  Kenyon, S.J. \& Hartmann, L., 1995, ApJS , 101, 117


\bibitem[Landolt (1992)]{land92} Landolt, A.U., 1992, \aj 104, 1

\bibitem[Matsuyama, Johnstone, \& Hartmann(2003)]{2003ApJ...582..893M} 
Matsuyama, I., Johnstone, D., \& Hartmann, L.\ 2003, \apj, 582, 893 

\bibitem[Meyer et al.(1997)]{mey97} Meyer, M., Calvet, N., Hillenbrand, L.A., 1997, \aj, 114, 288

\bibitem[Millan-Gabet et al.(1999)]{1999ApJ...513L.131M} Millan-Gabet, R., 
Schloerb, F.~P., Traub, W.~A., Malbet, F., Berger, J.~P., \& Bregman, 
J.~D.\ 1999, \apjl, 513, L131 

\bibitem[Miyake \&  Nakagawa 1995]{MN95}
Miyake, K., \& Nakagawa, Y. 1995, ApJ, 441, 361

\bibitem[Monnier \& Millan-Gabet(2002)]{2002ApJ...579..694M} Monnier, 
J.~D.~\& Millan-Gabet, R.\ 2002, \apj, 579, 694 

\bibitem[Muzerolle et al.(2000)]{muze00} Muzerolle, J., Calvet, N., Brice\~{n}o, C., Hartmann, L. \& 
Hillenbrand, L., 2000, \apj, 535, L47

\bibitem[Muzerolle et al.(2001)]{2001ysne.conf..245M} Muzerolle, J., 
Hillenbrand, L., Calvet, N., Hartmann, L., \& Brice{\~ n}o, C.\ 2001, ASP 
Conf.~Ser.~244: Young Stars Near Earth: Progress and Prospects, 245 

\bibitem[Muzerolle et al.(2004)]{muze04} Muzerolle, J., 
D'Alessio, P., Calvet, N., \&  Hartmann
2004, \apj, in press.


\bibitem[Natta et al.(2001)]{2001A&A...371..186N} Natta, A., Prusti, T., 
Neri, R., Wooden, D., Grinin, V.~P., \& Mannings, V.\ 2001, \aap, 371, 186 

\bibitem[Podosek \& Cassen(1994)]{podo94} Podosek, F.A. \&  Cassen, P., 1994, Meteoritics 29, 6-25

\bibitem[Pollack et al.(1996)]{poll96} Pollack, J., Hubickyj, O., Bodenheimer, P., Lissauer, J., Podolak, M.
 \& Greenzweig, Y., 1996, Icarus 124,62

\bibitem[Qi et al.(2003)]{2003ApJ...597..986Q} Qi, C., Kessler, J.~E., 
Koerner, D.~W., Sargent, A.~I., \& Blake, G.~A.\ 2003, \apj, 597, 986

\bibitem[Rice et al.(2003)]{2003MNRAS.342...79R} Rice, W.~K.~M., Wood, K.,
Armitage, P.~J., Whitney, B.~A., \& Bjorkman, J.~E.\ 2003, \mnras, 342, 79

\bibitem[Schlegel, Finkbeiner, \& Davis(1998)]{1998ApJ...500..525S} 
Schlegel, D.~J., Finkbeiner, D.~P., \& Davis, M.\ 1998, \apj, 500, 525

\bibitem[Siess et al.(2000)]{sies00} Siess, L., Dufour, E. \&  Forestini, M. 2000 A\&A , 358,593S

\bibitem[Tollestrup \& Willner(1998)]{1998SPIE.3354..502T} Tollestrup,
E.~V.~\& Willner, S.~P.\ 1998, \procspie, 3354, 502

\bibitem[Tuthill, Monnier, \& Danchi(2001)]{2001Natur.409.1012T} Tuthill,
P.~G., Monnier, J.~D., \& Danchi, W.~C.\ 2001, \nat, 409, 1012

\bibitem[Warren \& Hesser(1977)]{wah77} Warren, W.H., \& Hesser, J.E. 1977, \apjs, 34, 115
                                                                                
\bibitem[Warren \& Hesser(1978)]{wah78} Warren, W.H., \& Hesser, J.E. 1978, \apjs, 36, 497

\bibitem[Weidenschilling(1977)]{weid77} Weidenschilling, S.J., 1977, MNRAS, 180, 57

\bibitem[Weidenschilling \& Cuzzi(1993)]{weid93} Weidenschilling, S.J. \& Cuzzi, J.N., 1993, Protostars and
 Planets III, 1031

\bibitem[Wilner, Ho, Kastner, \& 
Rodr{\'{\i}}guez(2000)]{2000ApJ...534L.101W} Wilner, D.~J., Ho, P.~T.~P., 
Kastner, J.~H., \& Rodr{\'{\i}}guez, L.~F.\ 2000, \apjl, 534, L101

\bibitem[Wood (2004]{}
Wood, J. 2004. in The Search for Other Worlds (Eds. S. S. Holt, D. Deming), 14th Annual October Astrophysics Conference in Maryland (Oct. 13-14 2003, College Park MD), preprint at
http://cfa-www.harvard.edu/~jwood/


\end{thebibliography}
\end{document}